\documentclass[apj,twocolumn]{openjournal}
\usepackage{graphicx}
\usepackage{xcolor}
\usepackage{amsmath}
\usepackage{amsfonts}
\usepackage{amssymb}
\usepackage{upgreek}
\usepackage{txfonts}
\usepackage[utf8]{inputenc}
\usepackage{multirow}
\usepackage{booktabs}
\usepackage{longtable}
\usepackage{xurl}
\usepackage[pdfborder={0 0 0}]{hyperref}
\usepackage{cleveref}
\newcommand{\orcidauthor}[3]{\author{\href{http://orcid.org/#1}{#2 \openin1 orcid-ID.png \ifeof1 \else \hskip2pt\includegraphics[width=9pt]{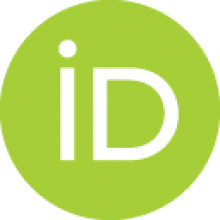}\fi}$^{#3}$}}

\def\frontmatter@title@below{\vspace*{0.1in}}
\makeatletter
\providecommand{\@spew@tblnotes}{}
\makeatother

\newcommand{\dm}{$\text{pc~cm}^{-3}$}

\newcommand{\tobs}{347.9}
\newcommand{\epochs}{154}
\newcommand{\GPn}{550,445}
\newcommand{\GPhundred}{5,474}
\newcommand{\cands}{1,048,138}

\DeclareUnicodeCharacter{2212}{-}
\DeclareUnicodeCharacter{2011}{-}

\graphicspath{{./}{figures/}}

\makeatletter
\AtBeginDocument{%
  \let\oaj@orig@sect\@sect
  \def\@sect#1#2#3#4#5#6[#7]#8{%
    \oaj@orig@sect{#1}{#2}{#3}{#4}{#5}{#6}[{#7}]{#8}%
    \cref@constructprefix{#1}{\cref@result}%
    \protected@edef\cref@currentlabel{[#1][\arabic{#1}][\cref@result]\csname p@#1\endcsname\csname the#1\endcsname}}%
}
\AtBeginDocument{\gdef\@arstrut@hook{\global\let\@arstrut\@arstrut@org}}
\let\oaj@lastpage@putlabel\lastpage@putlabel
\def\lastpage@putlabel{\oaj@lastpage@putlabel\global\let\lastpage@putlabel\relax}
\makeatother

\shorttitle{Crab DM Variations}
\shortauthors{Johnson et al.}

\begin{document}

\title{Over a Half-decade Study of Crab Giant Pulses Observed with I-LOFAR}

\orcidauthor{0000-0002-5927-0481}{Owen~A.~Johnson}{1,2,3,\dagger}
\orcidauthor{0000-0001-7185-1310}{David~J.~McKenna}{4}
\orcidauthor{0000-0003-4332-8201}{Sai~C.~Susarla}{5}
\orcidauthor{0000-0002-4553-655X}{Evan~F.~Keane}{1}
\orcidauthor{0009-0000-5589-0926}{Letizia~Vincetti}{1,6,7}
\orcidauthor{0000-0003-4399-2233}{Joseph~P.~McCauley}{1}
\orcidauthor{0000-0001-8208-4292}{Aaron~Golden}{8,9}

\affiliation{$^1$ School of Physics, Trinity College Dublin, College Green, Dublin 2, Ireland}
\affiliation{$^2$ Breakthrough Listen, University of California, Berkeley, 501 Campbell Hall 3411, Berkeley, CA, 94720, USA}
\affiliation{$^3$ Cahill Center for Astronomy and Astrophysics, California Institute of Technology, Pasadena, CA 91125, USA}
\affiliation{$^4$ ASTRON, The Netherlands Institute for Radio Astronomy, Oude Hoogeveensedijk 4, 7991 PD Dwingeloo, The Netherlands}
\affiliation{$^5$ School of Mathematical Sciences, University of Southampton, University Road, Southampton SO17 1BJ, UK}
\affiliation{$^6$ School of Cosmic Physics, Dublin Institute for Advanced Studies, 31 Fitzwilliam Place, Dublin 2, Ireland}
\affiliation{$^7$ Armagh Observatory and Planetarium, College Hill, Armagh BT61 9DG, N. Ireland}
\affiliation{$^8$ Physics, School of Natural Sciences \& Center for Astronomy, College of Science and Engineering, University of Galway, University Road, Galway, H91 TK33, Ireland}
\affiliation{$^9$ Department of Physics and Electronics, Rhodes University, Makhanda, 6140, South Africa}
\thanks{$^\dagger$ Corresponding author: \href{mailto:ojohnson@caltech.edu}{ojohnson@caltech.edu}}


\begin{abstract}
This work details the analysis of \cands~single pulses from the Crab pulsar detected with the Irish LOFAR station over a 6-year observation campaign spanning \epochs~epochs and \tobs~hours at 110-190~MHz. Of the detections in the sample \GPn~are giant pulses (GPs, with S/N$\geq10$), making it the largest analysed sample to date. The power law indices for the S/N, flux density and fluence distributions range from $3.0-3.4$ for the main pulse population and $3.3-3.5$ for the inter-pulse population. A subset of \GPhundred~GPs (with S/N$\geq 100$) were used to model scattering across the observing band to investigate the entanglement of dispersion measure (DM) and scattering. It was found that DM and scattering have a strong correlation coefficient of $0.70-0.71$ across the entire campaign. Removing the long-term trend using a running median windowed at $125–190$ days lowers the correlation to $0.48-0.50$ respectively. The implications of these DM and scattering observations from the perspective of both the Crab Nebula's structure and the solar wind are discussed, including a rise in DM from August to October 2024 indicating AU-scale density variation consistent with fine substructure within Rayleigh-Taylor shaped filamentary ejecta.

\end{abstract}


\maketitle


\section{Introduction}
The Crab pulsar (PSR~B0531$+$21) is one of the most notable and well-studied pulsars, owing to its well-determined, relatively young age and its exceptional brightness. Embedded at the centre of the Crab Nebula, the pulsar is the remnant of the historical supernova of 1054 CE with radio emission first noted by \citet{sr68}. The pulsar provides a unique opportunity to investigate both the physics of neutron stars and the dynamics of supernova remnants. The Crab emits strongly across the entire electromagnetic spectrum, making it an invaluable tool for studying the interaction between a pulsar and its surrounding Nebula \citep{hester_crab_2008}.

Due to its brightness and dynamic environment, the Crab pulsar is also an excellent source for studying emission and probing the free-electron content of the surrounding Nebula. At frequencies below 300 MHz, dispersion measure and scattering effects become particularly pronounced, and these effects have been observed consistently for decades in the Crab system~\citep{lyne_23_1993,ljg+15}. The evolving Nebula causes time-dependant propagation effects on monthly to yearly timescales \citep{bietenholz_new_2015}. An analysis of over half a decade of low-frequency observations of the Crab pulsar, conducted using the I-LOFAR radio telescope between 2020 and 2026, is presented. These observations provide a unique temporal baseline for investigating both the emission properties of the Crab pulsar and the evolving nature of the Crab Nebula along the line of sight over this period.

This work is structured as follows: \cref{sec:bkginfo} provides the scientific context and motivation for this study; \cref{sec:obsnmethod} describes the observations and methodology used for pulse detection and analysis; and \cref{sec:results_crab} presents a statistical overview of the giant-pulse population and examines DM variations and scattering timescales; Finally, \cref{sec:conclusion_crab} gives concluding remarks and potential future work on the topic. 

\section{Background \& Motivation} \label{sec:bkginfo}

\subsection{Giant Pulses}
The pulse amplitude distribution of the Crab pulsar is broad~\citep{hje15}, with the brightest pulses referred to as ``giant pulses'' (GPs). The working definition of GPs is inconsistent throughout the literature but for this work it is defined as those pulses that are brighter than ten standard deviations above the noise floor. GPs from the Crab pulsar are sporadic, their durations range from as long as milliseconds to as short as being unresolved on timescales of $\sim 2$~nanoseconds~\citep{jessner_giant_2010,hje15,thulasiram_narrow-band_2021}. GPs have also been noted to show narrowband features as a consequence of the Heisenberg-Gabor limit \citep{bij_kinematics_2021}. The peak flux density of a GP ranges from $10^{3}-10^{6}$ Jy~\citep{thulasiram_narrow-band_2021}, and they closely follow a power-law distribution of, $N(F)\propto F^\alpha$ where $\alpha$ ranges from $-1.5$ to $-3.6$, depending on pulse phase~\citep{karuppusamy_giant_2010,sokolowski_100_2025}. This indicates that these bright pulses are not exceedingly rare. The power-law distribution extends across at least three orders of magnitude in fluence, with no evidence of a high fluence cut-off. 

One of the most noteworthy things about GPs from the Crab pulsar is their brightness temperature. Single pulses can reach temperatures of at least $10^{32}~\rm K$ in observations that do not resolve the finest temporal structure, and are known to reach $10^{37}~\rm K$ in nanosecond-resolution observations \citep{cordes_brightest_2004}. This places GPs as some of the brightest signals in the observable Universe. Observation of giant pulses exhibit significant variability with detection rates measured from 100 -- 3000 pulses per hour. Intervals of lower and higher intensity activity show that successive detections exhibit a Poisson distribution with no evidence of `memory' \citep{sokolowski_100_2025}.

\subsection{The Crab Nebula}
The Crab Nebula has a stratified structure, at its core lies the Crab pulsar surrounded by relativistic magnetized wind flowing outward~\citep{lyubarsky_structure_2002}. This wind interacts with a termination shock, a boundary region where the wind's kinetic energy is converted into heated plasma and magnetic compression \citep{lyubarsky_structure_2002, gaensler_evolution_2006}. The shocked wind forms a synchrotron-emitting nebular interior composed of electrons, positrons, and wound-up magnetic fields \citep{hester_crab_2008}. Beyond this inner region, the synchrotron Nebula expands into supernova ejecta. This ejecta has cooled and condensed into regions made up primarily of ionized oxygen, forming a network of structures visible at optical wavelengths \citep{hester_crab_2008}. The filamentary structure exhibits both large-scale organization and small-scale turbulence, with the pulsar wind continuously pushing against these thermal structures creating a dynamic environment \citep{hester_crab_2008}.

The Crab Nebula contains a substantial population of free electrons distributed throughout all three regions, the wind, shocked nebular interior and the ejecta filaments \citep{kuzmin_deviation_2008}. Within the Nebula, the average electron number density is $\Delta n_e \sim 0.018~\rm{cm}^{-3}$ and this results in a dispersion measure contribution $\Delta \mathrm{DM} \sim 0.03$~\dm when using a Nebula radius of 1.7 pc \citep{kuzmin_deviation_2008,hester_crab_2008,mckee_temporal_2018}.

\subsection{Probing the Free Electron Content}


Because dispersive delay scales as $\Delta t\propto\text{DM}\nu^{-2}$ at low frequencies the delay is much more pronounced, making it easier to measure DM precisely. This makes telescopes like the LOw Frequency ARay \citep[LOFAR;][]{van_haarlem_lofar_2013} one of the best instruments for determining DM,  with long timing campaigns routinely achieving DM uncertainties on the order of $10^{-5}$~\dm \citep{iraci_pulsar_2024}. The exceptional brightness of GPs from the Crab Pulsar means that with a LOFAR international station residual smearing can be accounted for making GPs a highly sensitive probe of small DM offsets and short-timescale variations~\citep{susarla_long-term_2025}.

The Crab pulsar has a DM of 56.7712(3)~\dm according to the ATNF pulsar catalogue \citep{ATNFCat_2005}. However, this value is not static, as observations have demonstrated temporal variations on timescales of weeks to years on scales of $\pm0.01-0.15$~\dm~\citep{lyne_23_1993}. These variations indicate a change in electron density along the line of sight to the Crab pulsar and can be used as tracers for the surrounding Nebula \citep{isaacman_crab_1977}.


In addition to variability in the nebular electrons, the sun contributes to the observed DM from the Crab. The Sun continuously emits a magnetised plasma outflow, known as the solar wind, which fills the heliosphere with free electrons. Consequently, this can affect radio waves in any line of sight in the inner solar system \citep{tiburzi_impact_2021}. Any pulsar signal observed near the ecliptic plane can thus be impacted by the contribution of the solar wind, as is the case for the Crab pulsar. This is commonly modelled as $n\propto r^{-2}$, where $r$ is the distance from the sun, but in practice the structure also varies with heliographic latitude, longitude, and the solar cycle \citep{susarla_exploring_2024}. The contribution of the solar wind manifests in measurements by adding variations on scales of typically $10^{-4}$~\dm, with contributions an order of magnitude higher than this possible at and very close to solar conjunction. Thus, the solar wind contribution to the DM variability is typically sub-dominant to that of the turbulent nebula contribution in the case of the Crab. 
\section{Observations \& Methods} \label{sec:obsnmethod}

\subsection{Observations} 

\begin{figure*}[t]
    \centering
    \includegraphics[width=0.94\linewidth]{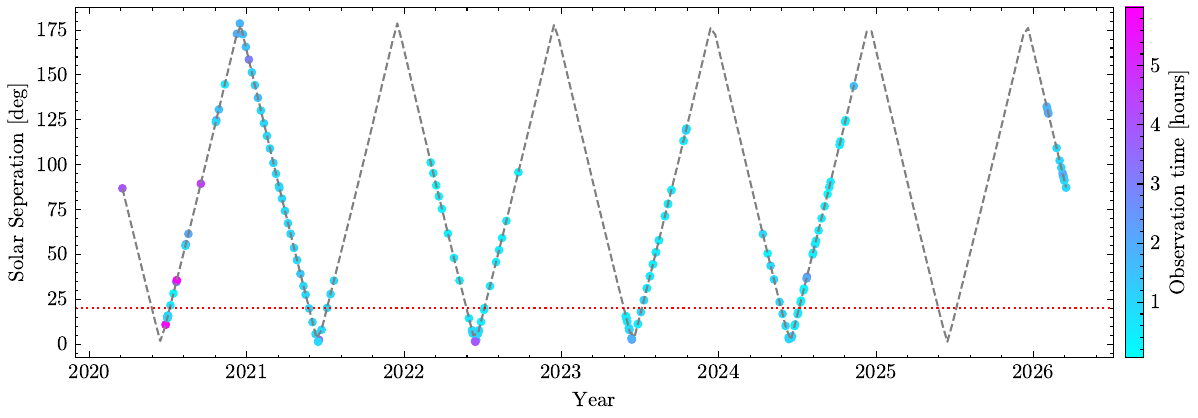}
    \caption{Each of the \epochs\ epochs covering a total of \tobs\ hours over 6 years of data plotted against solar separation.}
    \label{fig:crab_sun_seperation}
\end{figure*}

Observations were taken using the Irish LOw Frequency ARray station (I-LOFAR) using the REALTA backend \citep{murphy_first_2021} for processing over a time range from June 2020 to March 2026 totalling $\tobs$ hours and covering $\epochs$ epochs as shown in \cref{fig:crab_sun_seperation}. The high band antennas were used for each of the observations, which observed from $\sim 102.2-197.5$~MHz. During observations, UDP packets of beam-formed data are written to disk. The saved data are made up of coarse-channelized complex voltages.  The input data stream consists of two polarizations per antenna. After real Nyquist sampling with a 200-MHz clock and coarse channelization by a factor of 512, only 488 channels (or 244 channels) are recordable when the data is written as 8-bit (or 16-bit) complex samples; for this study 8-bit mode is used. The \textsc{udpPacketManager} \citep{David_JOSS} is then used to create \textsc{sigproc} \citep{lorimer_sigproc_2011} formatted filterbanks at the specifications detailed in \cref{tab:data_details_crab}

\begin{table}[htbp]
    \centering

    \begin{tabular}{lc}
        \hline
        \textbf{Data Attribute} & \textbf{Value} \\
        \hline
        Frequency Start $(f_\text{start})$ & 102.234131 MHz \\
        Frequency End $(f_\text{end})$ & 197.546387 MHz \\
        Frequency Resolution & $24.4414~\mathrm{kHz}$ \\
        Temporal Resolution & $655.36~\upmu\mathrm{s}$ \\
        \hline
    \end{tabular}
    \caption{Specifications of data generated with \textsc{udpPacketManager}~\citep{David_JOSS}.}
    \label{tab:data_details_crab}
\end{table}

\subsection{Identifying Pulses}

\begin{figure*}
    \centering
    \includegraphics[width=0.8\linewidth]{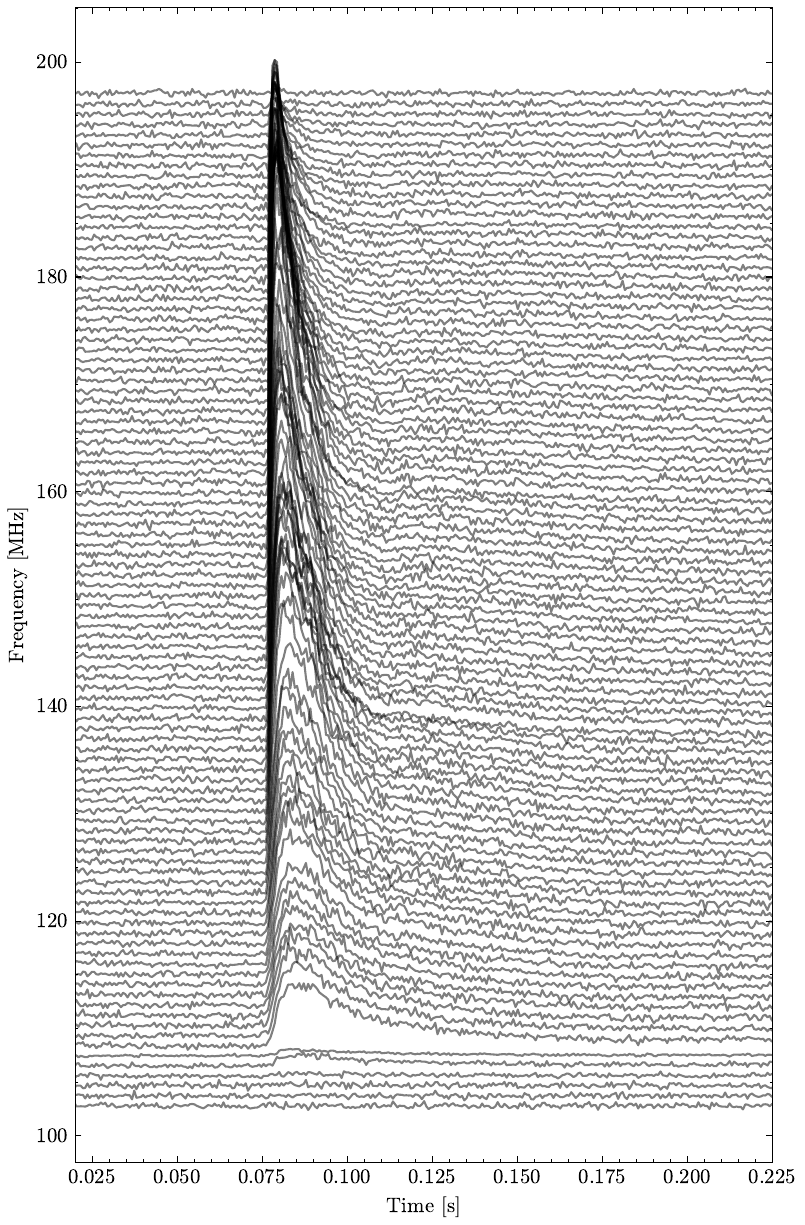}
    \caption{Waterfall plot of the brightest giant pulse across all epochs. The burst was observed at 2023-10-17T04:33:38.896 with a corrected DM of $56.754\pm0.002$~\dm and an S/N of 7214. A hardware filter is used for the FM band; the roll-off from this filter is evident at the lowest frequencies.}
    \label{fig:joy_division}
\end{figure*}

The first step in analysing these data is the identification of bright single pulses; this is done using \textsc{transientX}~\citep{men_transientx_2024}. 
Single pulses with widths of $5-50$~ms were searched for to generate candidate lists for each epoch, 5~ms was chosen due to ISM scattering at the low end of the band being $\sim6.32~\mathrm{ms}$ \citep{bhat_clean-based_2003,ocker_ne2025_2026}. After an initial pass, the pulses were reprocessed with \textsc{transientX} a second time with a finer DM grid to maxmise the S/N of each pulse. To reduce computational cost, the search was performed on 8-bit filterbanks rather than 32-bit. The candidate lists provide time-stamps used to extract portions of the 32-bit filterbanks. For each burst where an S/N of $>100$ was measured, a \textsc{psrchive} \citep{hotan_psrchive_2004} single-pulse archive file was generated, where $t_{\rm start} =  t_{\rm burst}- 3.276~\rm ms$ and $t_{\rm end} =  t_{\rm burst}+ 32.768~\rm ms$. This window was chosen such that the single-pulse filterbanks didn't include multiple rotations and the sample start time was 10\% of a rotation before the burst recorded burst time. The brightest pulse in the dataset was found to have an S/N of 7214 and is shown in \cref{fig:joy_division}, this is a prime example of the benefit of observing GPs at low frequency as the scattering is easily observed in each of the 0.95-MHz sub-bands.

\subsection{Correcting the DM}

\begin{figure*}
    \centering
    \includegraphics[width=0.8\linewidth]{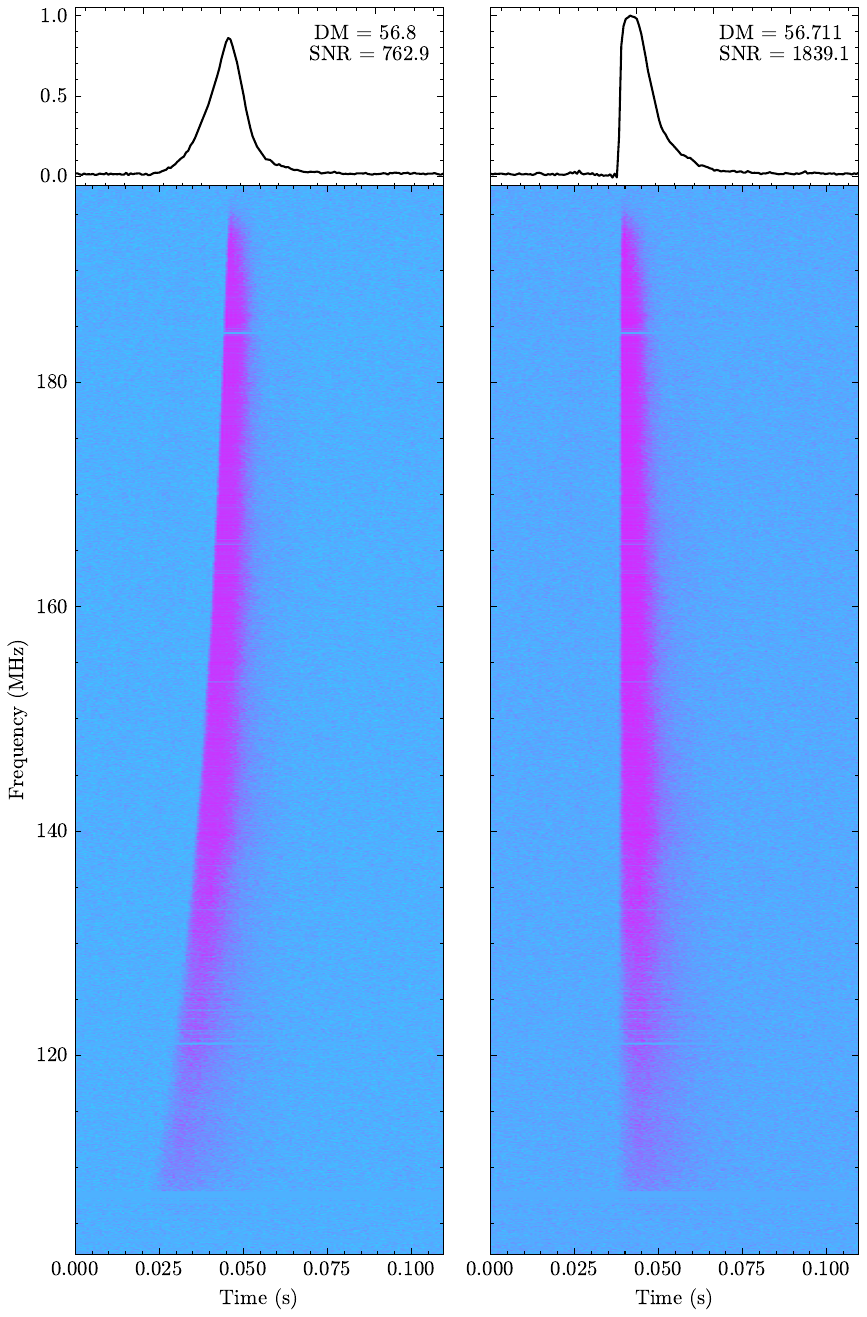}
    \caption{The \textit{left} shows an example of a giant pulse where the initial DM trial that \textsc{transientX} determined by fitting a boxcar, the \textit{right} shows the phase corrected DM that was calculated by \textsc{DM\_phase}}
    \label{fig:DMvariation}
\end{figure*}

Due to the large amount of scattering present at lower frequencies, traditional boxcar single-pulse searches can often over or underestimate the DM as they simply maximise S/N over a reasonably coarse search grid in DM and pulse width. An example of this is shown in \cref{fig:DMvariation}. With this in mind the initial DM, from the \textsc{transientx} detection was corrected by using \textsc{DM\_phase} \citep{seymour_dm_phase_2019}, which optimises based on coherent radio power (phase of the pulse) across the band. This is done by ingesting RFI-cleaned archive files and dedispersing across a range of trial DMs. The trial DM that maximises inter-channel phase coherence, rather than the amplitude of the dedispersed time series, is selected. The corresponding dispersive delay, $\Delta t$ is applied. After dedispersion, the intensity waterfall $I(\nu_i,t)$ is Fourier transformed along the pulse phase/time axis to obtain $\tilde{I}_i(f)=\mathcal{F}_t\{I(\nu_i,t)\}$. The phase-only spectrum is then formed by normalising the amplitude like:
\begin{equation}
\hat{I}_i(f)=\frac{\tilde{I}_i(f)}{\left|\tilde{I}_i(f)\right|}~.
\end{equation}
This is then summed coherently across all frequency channels like:
\begin{equation}
C(f;\mathrm{DM})=\sum_{i=1}^{N_\nu}\hat{I}_i(f), \qquad
P(f;\mathrm{DM})=\left|C(f;\mathrm{DM})\right|^2 ~.
\end{equation}
A one-dimensional DM metric is obtained by integrating $P(f;\mathrm{DM})$ over a range in fluctuation frequency $f$,
\begin{equation}
S(\mathrm{DM})=\sum_{f=f_{\rm low}}^{f_{\rm high}} w(f)~P(f;\mathrm{DM})~,
\end{equation}
and the best-fit DM is selected as the maximum $S(\text{DM})$. The DM uncertainty is estimated by fitting a polynomial to the peak of $S(\mathrm{DM})$ and using the curvature at the maximum (Hessian approximation) $\mathrm{DM}_0$, giving
\begin{equation}
\sigma_{\mathrm{DM}}=\sqrt{\left|\frac{2~\varepsilon}{{S_{\text{max}}}''(\mathrm{DM}_0)}\right|}~ ,
\end{equation}
where $\varepsilon$ is a noise scale for the DM metric.

\subsection{Modelling Scattering} \label{sec:modelScattering}


Given the precision with which DM can be determined from a single international LOFAR station \citep[pp.~208--220]{david_thesis}, it is possible to model the scattering of individual pulses as a function of frequency. Splitting pulses into 10-MHz chunks, three models have been previously used to model scattering as a function of frequency and time \citep{bhat_detection_2007, karuppusamy_giant_2010, sokolowski_100_2025}. 
The thin screen model proposed by \cite{bhat_clean-based_2003}, assumes that scattering occurs within a narrow layer located on an infinite plane. Because the screen is sufficiently thin, radio waves crossing it experience no change in propagation speed, and the observed delay arises purely from the geometric path difference caused by the scattering event. It is described as follows: 

\begin{equation}
    f_{\text{thin}}(t, t_0,\tau,A) = \frac{A}{t} \exp \left[ -\frac{t - t_0}{\tau} \right] H (t - t_0)
    \label{eq:thinmodel}.
\end{equation}

This models the pulse with an exponential decay with a characteristic value for scattering, $\tau$. Here $H(t-t_0)$ is a delayed Heaviside step function for an event that occurs at $t_0$. It switches the model on at $t>t_0$ and off for $t<t_0$. $A$ is then the amplitude used to scale to the pulse profiles if they are not normalised. Observationally this model did not fit at lower frequencies and so a modified version of the thin model was implemented by \cite{karuppusamy_giant_2010}. This modification added a scaling factor of $t^\gamma$ to better emulate the observed rise of the exponential peak. This is described by \cref{eq:thin_model_mod}, 

\begin{equation}
    f_{\text{mod, thin}} (t_, t_0, \tau, A, \gamma)= A(t-t_0)^\gamma \exp \left[ -\frac{t - t_0}{\tau} \right] H(t-t_0). 
    \label{eq:thin_model_mod}
\end{equation}

Similarly \cite{williamson_pulse_1972} describes a thick model. In the thick model, radio waves are continuously scattered during their propagation from the source to observer through a fraction of the medium. In this case, \cref{eq:thick_model} places the screen close to the source, which can be used to model the effect of the Crab's nebula on individual pulses.

\begin{equation}
    f_{\text{thick}} (t,t_0,\tau,A) = \sqrt{\frac{A \pi \tau}{4(t-t_0)^3}} \exp\left[ - \frac{\pi^2 \tau}{16(t-t_0)} \right] H(t -t_0)
    \label{eq:thick_model}
\end{equation}

\begin{figure}
    \centering
    \includegraphics[width=0.8\linewidth]{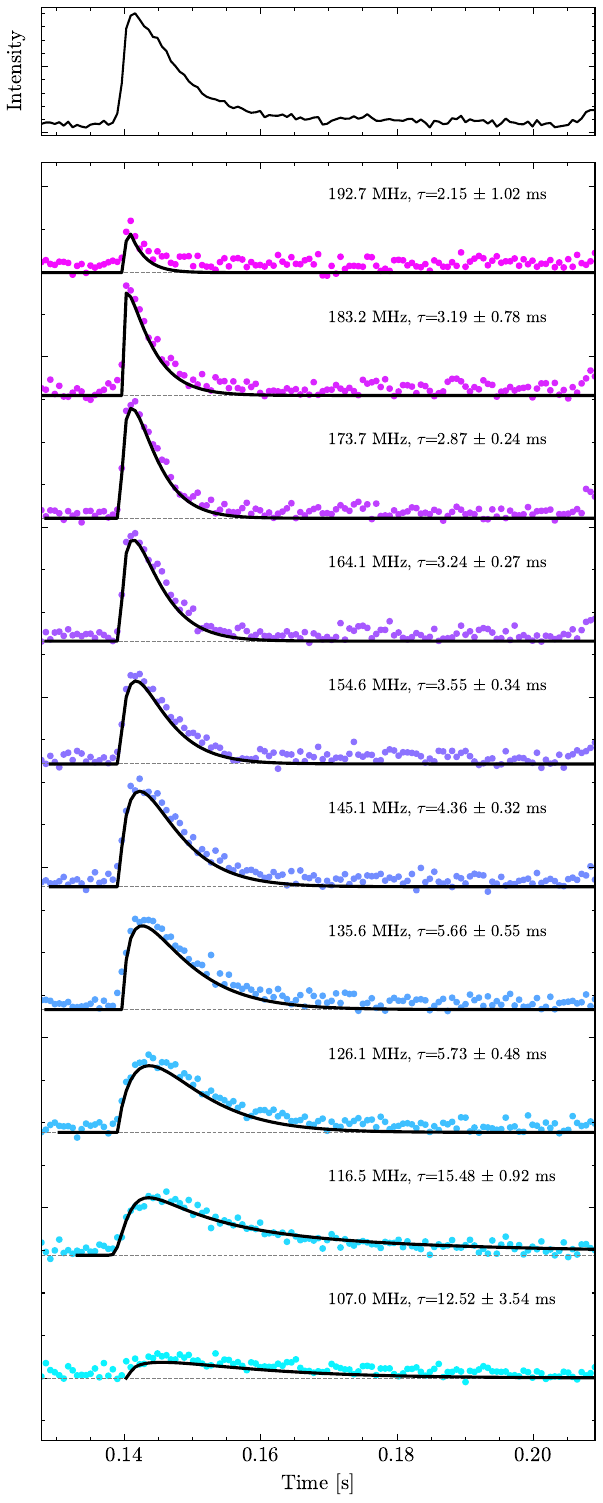}
    \caption{Example of the scattering fits applied to an example giant pulse that has a post-corrected S/N of 291 which is split into 9.5~MHz subbands. Here the black line represents the fitted scattering model; the scatter points are the observed data. }
    \label{fig:scattering}
\end{figure}

For each of the Crab GPs that exceeds 100~S/N and have been DM-corrected, the respective archive file is read and split into 9.53-MHz subbands (see \cref{fig:scattering}). Each subband is then fitted to both, \cref{eq:thin_model_mod} and \cref{eq:thick_model} using \textsc{lmfit} \citep{newville_2025_16175987} to compute statistical significance of each fit in a number of ways. In this case, Akaike Information Criterion (AIC) is used to determine which model fit was optimal for each subband. AIC was chosen as it quantifies the information loss between models~\citep{akaike_new_1974} with the lowest scoring model being marked as the preferred in that specific subband for that burst. \Cref{eq:thinmodel} was omitted from fitting due to previous studies \citep{karuppusamy_giant_2010} and preliminary work on this data set~\citep{david_thesis} where the model failed to converge for any of the observed subbands in the frequency range of this study. 

\subsection{Folding}

\begin{figure}
    \centering
    \includegraphics[width=0.8\linewidth]{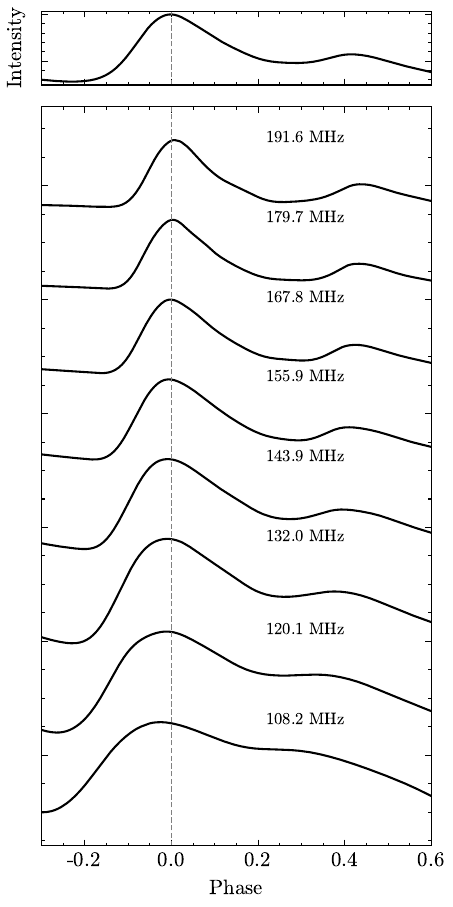}
    \caption{Smoothed folded profile of a 2-hour observation taken on 2026-03-17. Split into 12~MHz chunks.}
    \label{fig:foldedprofile}
\end{figure}

In addition to the single pulse analysis, each of the epochs were also folded for a comparison to the DMs measured from the Crab GPs. For each epoch, the Crab ephemeris measured monthly by Jodrell bank was used\footnote{\url{https://www.jb.man.ac.uk/pulsar/crab/}} \citep{lyne_23_1993} as a first estimate, \textsc{PRESTO} \citep{ransom_presto_2011} was then used to further refine DM and period values before timing with \textsc{TEMPO2} \citep{hobbs_tempo2_2006}. Initial efforts were made to time each epoch as described in \citet{iraci_pulsar_2024} and \citet{susarla_long-term_2025} which is briefly outlined below. However, determining an accurate ephemeris for the Crab across over half a decade proved difficult due to scattering exhibited in the Crab's folded pulse profile, along with DM variations on short timescales (see. \cref{fig:foldedprofile}), as pulsar timing relies on the ability to lock onto sharp features. DM was determined using single epoch timing values through the creation of a template from a high S/N observation and then fitting DM to from Times of Arrival (TOAs) for 8 subbands in the archive. Due primarly to scattering the uncertaintty was on the order of $10^{-2}$~\dm for each of the epochs, whereas for other pulsars not embedded in a complex Nebula it is on the order of $10^{-5}$~\dm~\citep{susarla_long-term_2025} with the same observing system. The singe-pulse DM precision is thus much higher in the case of our study. As a result, the folded analyses were omitted from our DM time series (below) as they provided no further insights beyond their higher frequency contemporaries (see fig. 3 of \citealt{karuppusamy_crab_2012})

\section{Results} \label{sec:results_crab}

This work details the analysis of $\tobs$ hours of Crab data over 6 years using the I-LOFAR telescope. In this dataset $\GPn$ giant pulses were detected in the dataset. Of these GPs $\GPhundred$ had S/N greater than 100 and were used for the finest DM measurements and for the modelling scattering. 

\subsection{Giant Pulse Sensitivity Calculations}

Overall $\cands$ pulses were detected above the average noise floor of 7.3 S/N, of these $\GPn$ were GPs. Generating a histogram of just the GPs and fitting a power law to this population of pulses is shown in \cref{fig:SNR_histogram}. Calculating the flux distribution was done using a modified version of the radiometer equation as described in \citet{johnson_radio_2025} and \citet{vincetti_characterising_2026}: 

\begin{equation}
\begin{split}
S_{\min,\rm{chan}}(f, l, b) = {} & \frac{\left[ T_{\text{sky}}(l,b) + T_{\text{int}}(f) \right] \cdot 2 k_B}{A_{\text{eff}}(f) \cdot m_{\text{II}} (f, l,b)} \cdot \beta \\
& \times \frac{\text{S/N}_{\min} }{\sqrt{n_p~ \tau_{\mathrm{burst}}~ B_{\text{channel}}}} \cdot 
\end{split}
 \label{eq:radiometer_equation}
\end{equation}

Here, $T_{\text{sky}}(l,b)$ is the sky temperature at Galactic coordinates $(l,b)$, $T_{\text{int}}(f)$ is the frequency-dependent instrumental temperature, $k_B$ is the Boltzmann constant, $A_{\text{eff}}(f)$ is the effective collecting area of the telescope, and $m_{\text{II}}(f,l,b)$ is the beam-model sensitivity correction factor. The factor $\beta$ accounts for instrumental losses, $\text{S/N}_{\min}$ is the minimum detectable signal-to-noise ratio, $n_p$ is the number of summed polarizations, and $B_{\text{channel}}$ is the channel bandwidth. The final term $\tau_{\mathrm{burst}}$ is the width of the pulse, but this can also be replaced with $t_{\rm obs}$ for observation duration and add a unitless factor of $\sqrt{W/(P-W)}$ which accounts for pulse broadening, where $P$ is the pulse period and $W$ is the observed pulse width, this is used in cases where the observations are folded. 

In the case of $T_{\rm sys}$ and $T_{\mathrm{int}}$ the values are calculated as described in \cite{mckenna_census_2024}, values for $A_{\mathrm{eff, max}}$ are taken from \cite{van_haarlem_lofar_2013}. In the case of the $m_{\mathrm{II}}$, the values are taken as the average over a four-hour track of the Crab using I-LOFAR, peaking at $59^{\circ}$ in elevation with the same method as described in \cite{vincetti_characterising_2026}. This was chosen as observations are typically taken very close to max elevation. 

The outputted value from \cref{eq:radiometer_equation} is the average flux density measurement for a given channel in Jy, so the combined flux density measurement across the band is calculated using quadrature summation of each channel as follows: 

\begin{equation}
    S_{\mathrm{total}} (f,l,b) = \left[ \sum_f \frac{1}{S_{\mathrm{chan}}(f,l,b)^2} \right]^{-1/2}
\end{equation}

The details of the population of observed bursts from each day are shown in table \ref{tab:bursts}. Note that the burst rate decreases with angular separation from the zenith, consistent with the decreased effective area. 




\subsection{Pulse Distributions}

\Cref{fig:SNR_histogram} shows the distributions of pulses across phase, S/N and flux density and fluence. Fluence, $\mathcal{F}$, was calculated for each pulse detected above the noise floor. The fluence calculation used the determined boxcar width, essentially an equivalent width, and a peak flux density according to the S/N obtained from \textsc{TransientX} with this width, such that $\mathcal{F} = {W} \cdot S_{\text{burst}}$. The resultant distribution when first fitting the sample shows an excess towards the tail of the distribution. This excess is more pronounced in the flux denisty and fluence distribution but present in all three. Initially, this was thought to be the contribution of two different populations with two different representative power-laws, the main pulse (MP) and the inter-pulse (IP). the MP is the dominant emission component and the IP is the secondary peak in the pulse profile shown in \cref{fig:foldedprofile} and \cref{fig:SNR_histogram} (a). The IP is usually treated as a distinct magnetospheric emission component that peaks around 0.9 in phase. It is thought to be linked to the opposite pole or outer magnetosphere, although its high-frequency behaviour suggests a more complex geometry than the standard two-pole picture \citep{cordes_brightest_2004, petrova_physics_2008}. 

Because of this, each burst was separated, categorised as MP or IP based on the calculated phase. Due to scattering the exact point in phase where to delineate between IPs and MPs is difficult to determine. \cite{driessen_scattering_2019} investigated unique pulse morphologies based on variations in phase location and varying rise and decay properties caused by scattering, yielding a set terminology for these features \citep[e.g. fig 7;][]{driessen_scattering_2019}. Two features of note from \citet{driessen_scattering_2019} are `echos' which are delayed replicas of the Crab pulsar's profile with a truncated exponential shape which form from `bumps' which are smoother approximately Gaussian delayed components that approach and recede between the MP and IP over time. However, none of these features were present in the sample of GPs identified in this work that were fitted with scattering models. This is likely due to any `echo' or `bump' being adsorbed into the broadened pulse which likely contributes to the overall pulse broadening observed. 

\begin{figure}
    \centering
    \includegraphics[width=0.9\linewidth]{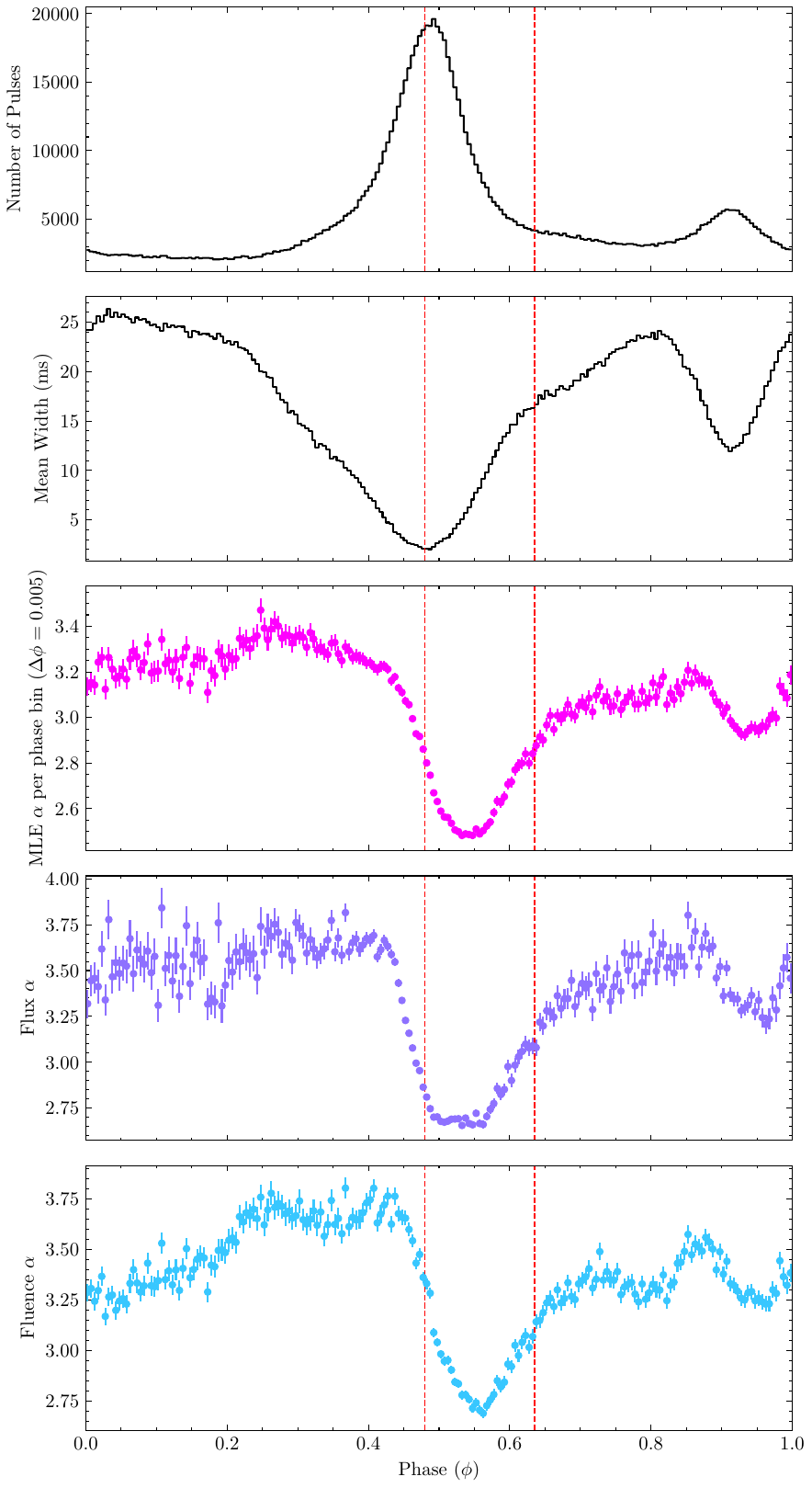}
    \caption{The top panel shows the number of pulses per phase bin of width 0.005. The bottom panel shows the MLE $\alpha$ value calculated for each phase bin. The dotted lines represent the best log-likelihood MP phase range.}
    \label{fig:MLEphase}
\end{figure}

\begin{figure*}[htbp]
    \centering
        \begin{minipage}{0.48\linewidth}
        \centering
        \includegraphics[width=\linewidth]{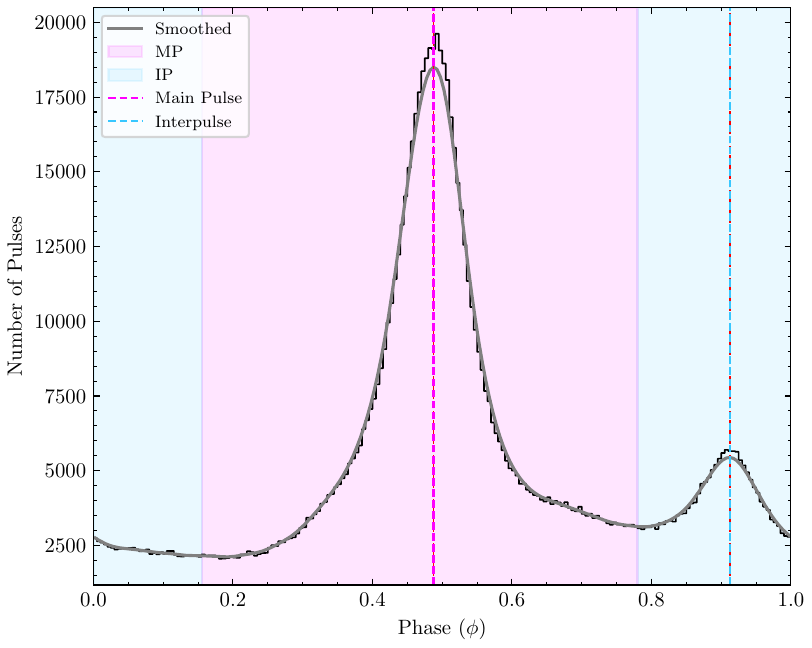}
    \end{minipage}
    \hfill
    \begin{minipage}{0.48\linewidth}
        \centering
        \includegraphics[width=\linewidth]{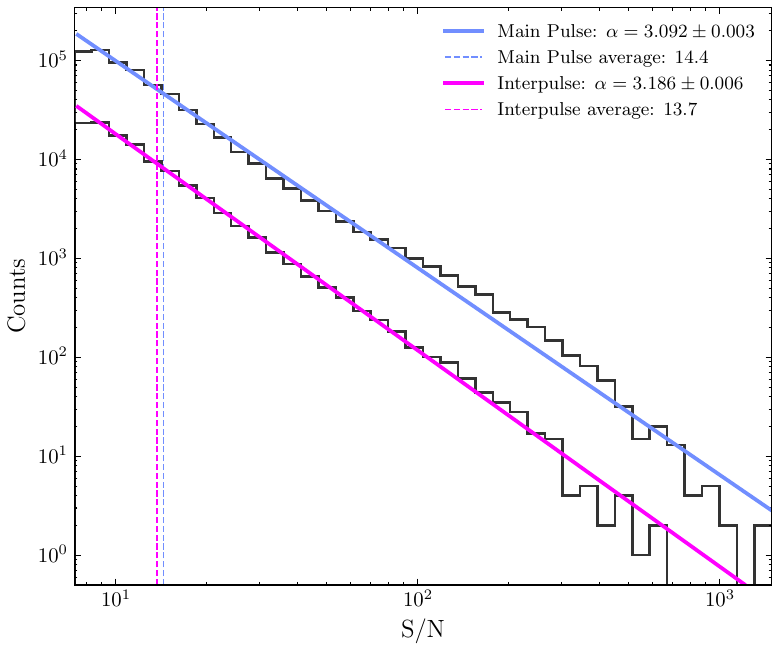}
    \end{minipage}
    \vspace{0.5cm}
    \hfill
    \begin{minipage}{0.48\linewidth}
        \centering
        \includegraphics[width=\linewidth]{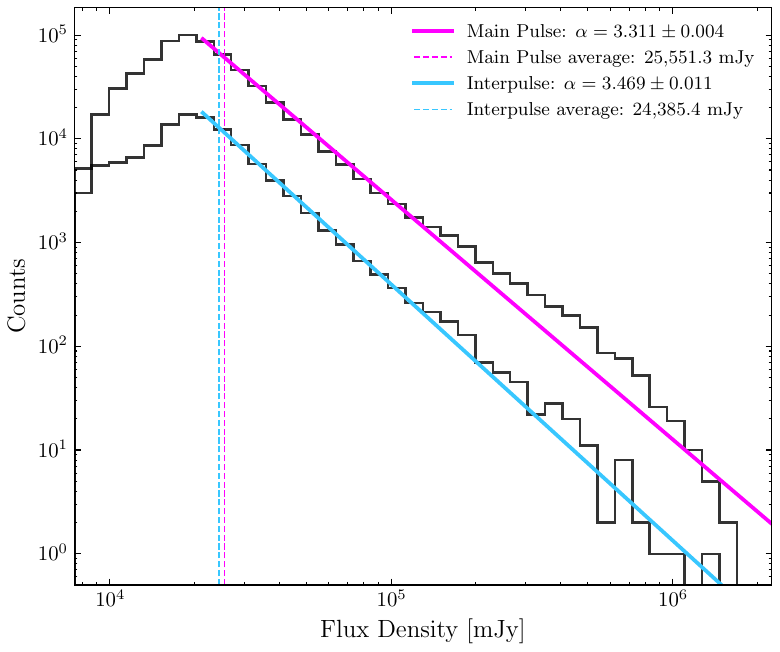}
    \end{minipage}
    \begin{minipage}{0.48\linewidth}
        \centering
        \includegraphics[width=\linewidth]{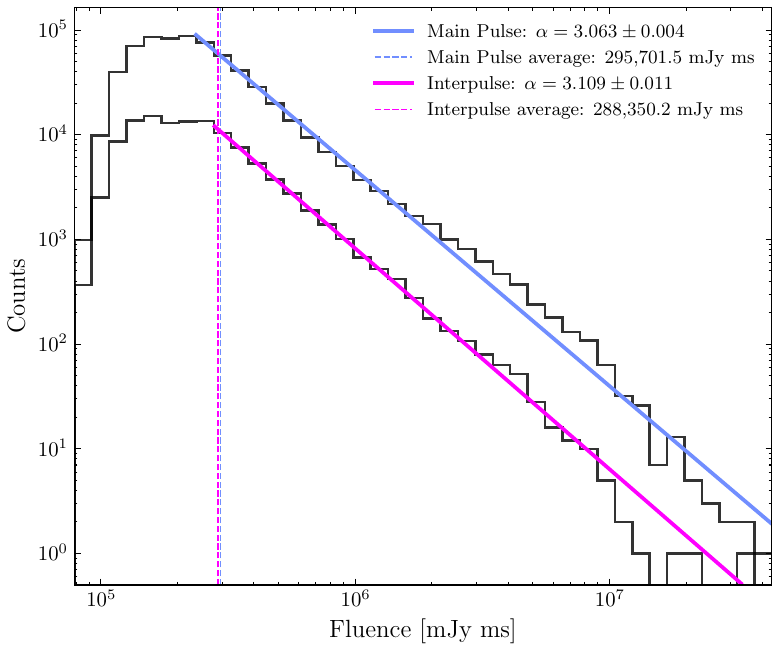}
    \end{minipage}

    \caption{Properties of the detected Crab pulses.\textit{a)}:  pulse phase distribution. \textit{b)}: signal-to-noise ratio distribution. \textit{c):} flux density distribution. \textit{d):} fluence distribution.}
    \label{fig:SNR_histogram}
\end{figure*}

To try and estimate the phase boundaries of the two pulse population a maximum likelihood estimation \citep[MLE;][]{clauset_power-law_2009} is used. This is done by binning each of the S/N values into phase bins of width 0.005 where a power law $(Ax^{-\alpha})$ fit is computed, shown in \cref{fig:MLEphase}. This shows two distinct features, one at $\phi = 0.45-0.7$ where $\alpha=2.7$ and a much weaker feature at $\phi = 0.9-1.005$ with a $\alpha=3.3$. These phase ranges $\alpha$ values are consistent with other studies \citep[i.e.][]{karuppusamy_crab_2012} where the MP population is considered to occur at $\phi \sim0.5$ and have a flatter component than that of the IP population which is noted to occur at $\phi \sim 0.9$. The error for each $\alpha$ value is calculated as follows: 

\begin{equation}
    \Delta\alpha = \frac{(\alpha - 1)}{\sqrt{n}}.
\end{equation}

This error estimation is purely statistical and assumes that each the distribution of S/N values in a phase bin can be represented by a power-law \citep{clauset_power-law_2009}. The two features in \cref{fig:MLEphase} indicate that there are two populations as expected, but the presence of a baseline $\alpha$ of 3.6 implies that there are GPs that may have incorrect phase assignment due to scattering or other factors as mentioned. 

Assuming that there are only two populations within the dataset an estimation of the phase windows of MPs and IPs can be made. A sliding window of varying phase length ($L_{\phi}: 0.005 -0.5$) and start phase ($\phi_0  : 0.005-0.995$) is used to represent the MP population with anything outside this window representing the IP population. Each population for a given window has a separate MLE power law fitted the same as above with the maximum log-likelihood values taken for each fit. These two likelihoods are summed to give that given window a `total' likelihood metric, with the window with the highest likelihood selected as the best phase ranges. This best windowed result is then compared to an overall fit of the entire population using Bayesian information criterion to ensure that the two population fit is preferred oppose to a single population. This method fitted the main MLE feature, as shown in \cref{fig:MLEphase} by the dashed vertical lines, however, this is inconsistent with the physical interpretation shown by the pulse profile (top panel fig. \ref{fig:MLEphase}). The determined window shows a one-sided fit that begins near the peak of the phase plot where MPs are seen at higher frequencies \citep[i.e.][]{lyne_23_1993} and ends before the scattering tail from the MP plateaus.  

As a result, the population window was determined by fitting a 1D Gaussian filter\footnote{\url{https://docs.scipy.org/doc/scipy/reference/generated/scipy.ndimage.gaussian_filter1d.html}} to the phase profile and taking the points of inflection of the two peaks as the `start' point for MPs and IPs which was determined as 0.487 and 0.912 which is consistent with the literature.  The end of the window was then taken to be an estimation of the local minimum in the wells between the peaks of the phase profile, this was determined by using the middle of the 10\% lowest values in each well. MPs were considered the population that fell into the pink region in \cref{fig:SNR_histogram} (a) and IPs anywhere outside this region, marked in blue. The excess remained after the populations had been separated out (shown in \cref{fig:SNR_histogram}) with their respective fitted power law between the MP and IP populations, with IP population being steeper in the case of all three. 

These values are in-line with what is expected, the MP population has been well noted to be intrinsically brighter than the IP population. As discussed the separation of these two populations is difficult. The more pronounced excess in the fluence distribution is likely due to the higher S/N pulses introducing a bias as the fitted scattering timescale is easier to discern from the noise floor causing pulses with a higher S/N to contribute the more representative fluence and flux density values in our sample, as opposed to the low S/N population. Overall the current results are therefore best interpreted as suggestive, but not as a definitive measurement of intrinsic pulse-phase or fluence boundaries between IP and MP population and the Crab as a whole, with the measured slope per phase in \cref{fig:MLEphase} providing a more reasonable estimate as it has a more reliable segregation of the two pulse populations.

\begin{figure}
    \centering
    \includegraphics[width=0.96\linewidth]{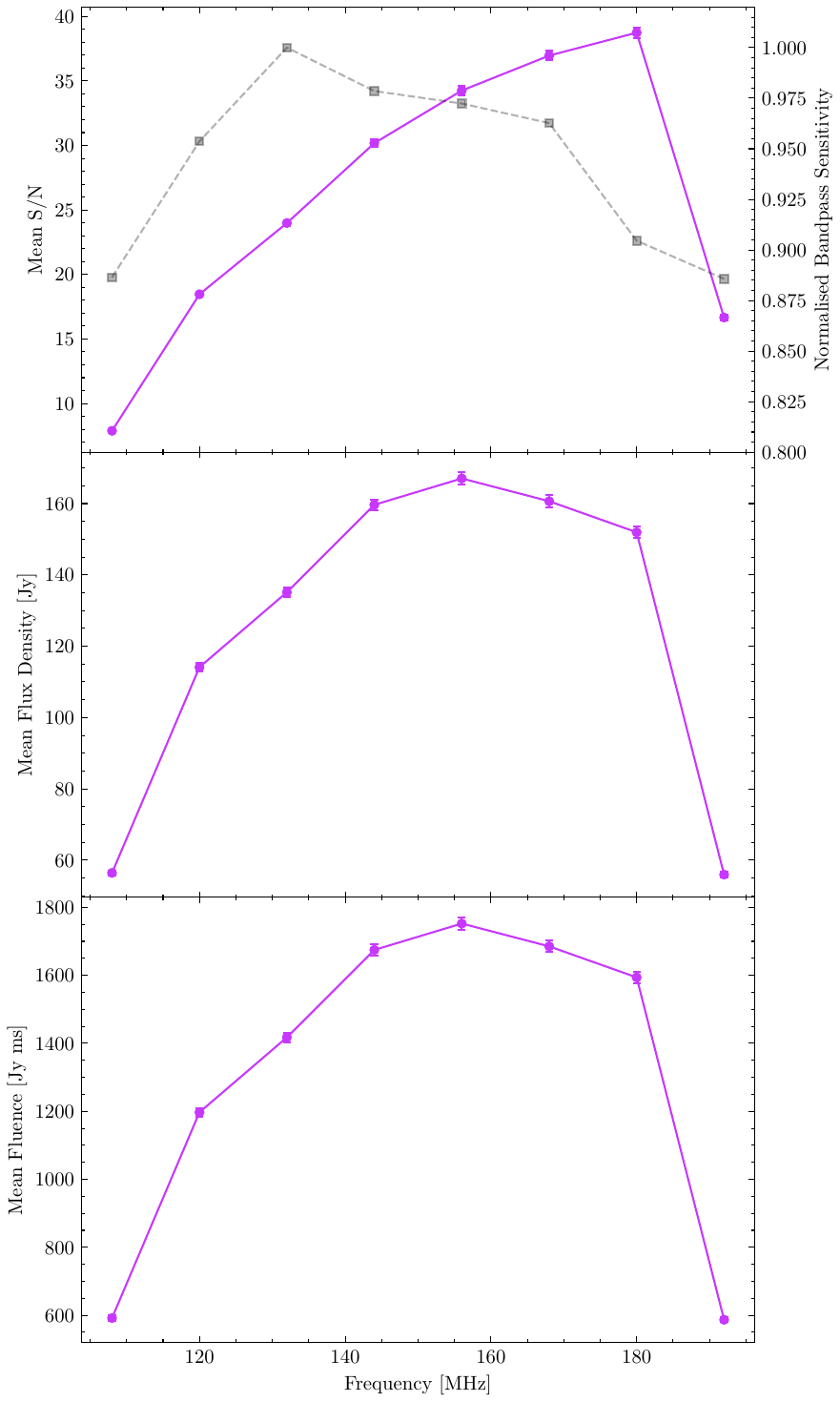}
    \caption{Frequency as a function of S/N \textit{(left)}, mean flux density \textit{(middle)} and mean fluence \textit{(right)} for GPs with an S/N over 100. The normalised bandpass response is shown in grey.} 
    \label{fig:GPspectra}
\end{figure}

The fitted power‑law indices for S/N, flux density and fluence in our Crab giant‑pulse sample (MP $\sim3.0–3.4$; IP $\sim3.3–3.5$) fall on the upper end of previously published ranges for Crab GPs, which typically report indices between $1.6–3.6$ at these frequencies depending on the chosen observable \citep{karuppusamy_giant_2010, karuppusamy_crab_2012, sokolowski_100_2025}. The spread in fitted power-law indices reflects differences in observing frequency, fit range, completeness, and scattering bias which has been showcased here. As a result, there is no universally accepted value for each population but rather an expected range based on frequency, chosen subband and so on.  


\Cref{fig:GPspectra} shows the S/N, flux density and fluence spectra averaged over the \GPhundred~GPs with S/N exceeding 100. The average S/N spectrum broadly follows the normalised bandpass response of I-LOFAR \citep{van_haarlem_lofar_2013}, reflecting the reduced instrumental sensitivity towards the edges of the observing band. At lower frequencies, the S/N is further diminished by increased scattering, an effect that is also evident in the flux density and fluence spectra.

The S/N spectra of the \GPhundred~GPs with S/N exceeding 100 broadly follow the normalised bandpass response of I-LOFAR \citep{van_haarlem_lofar_2013}, reflecting the reduced instrumental sensitivity towards the edges of the observing band. At lower frequencies, the S/N is further diminished by increased scattering, an effect that is also evident in the flux density and fluence spectra.








\subsection{Scattering}

As previously discussed scattering was fitted to each of the GPs above an S/N of 100 after using \textsc{DM\_phase}. In observations where no GP greater than 100 was observed, the next brightest pulse was used. As described in §\ref{sec:modelScattering} both modified thin (eq. \ref{eq:thin_model_mod}) and thick (eq. \ref{eq:thick_model}) models were fitted to the data and their statistical significance noted using \textsc{lmfit}. The scattering timescale in each sub-band of each pulse was measured, the median scattering value was taken for both the thin and thick scattering models.  A power law in the following form

\begin{equation}
    \tau(f) = \tau_{\mathrm{150}} \left[\frac{f}{150~\text{MHz}} \right]^\alpha\;,
    \label{eq:powerlawscattering}
\end{equation}

was then fit to the median sub-band values using non-linear least-squares optimisation. Here $\tau_{150}$ is the scattering timescale scaled to 150~MHz and $\alpha$ is the scattering index. The resultant values for each epoch are shown in \cref{fig:scattering} where the errors in this case are estimated from the covariance matrix of each fit. 


The lower the observing frequency, the smaller the DM uncertainty, since the dispersive delay across a fixed bandwidth scales as $\sim f^{-2}$, in theory improving the accuracy of the determined DM. However, at these frequencies scattering is also substantially increased $\sim f^{-4}$, leading to large broadening in the observed pulse, as observed in \cref{fig:scattering}. Consequently, a trade-off arises in the ability to measure DM and the increased temporal offset that comes with scattering.  

Calculating the `raw' correlation between both the scattering models and measured DM shows a positive correlation. The thin model has a Pearson score, $r$, of $0.71 \pm 0.04$ and a Spearman score, $\rho$, of $0.78 \pm 0.04$, with the thick model having $r = 0.70 \pm 0.04$ and $\rho = 0.75 \pm 0.04$ respectively. The higher Spearman values in both cases suggest that the relationship is monotonic but not linear. The correlation here is higher to that found by \cite{mckee_temporal_2018} which was found to be $0.56\pm 0.1$. DM and scattering both drift with time and no neighbouring days were observed to be independent which is in-line with what's expected in this physical picture. 





\begin{figure}
    \centering
    \includegraphics[width=0.95\linewidth]{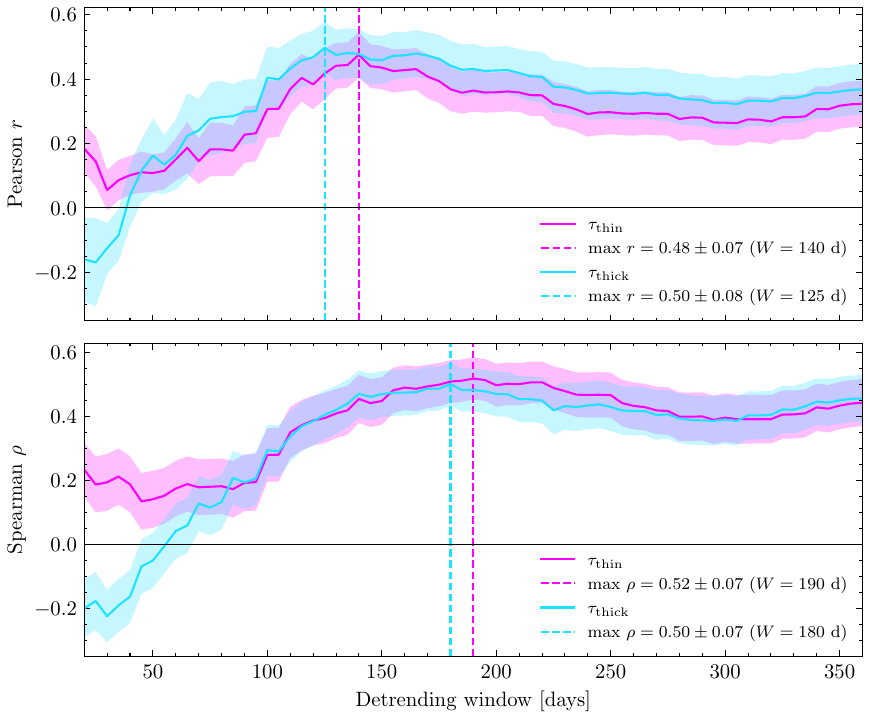}
    \caption{De-trended correlation of DM and scattering for different running median windows.}
    \label{fig:detrended_correlation}
\end{figure}

To remove the long-term correlation a similar method to both \citet{lam_systematic_2016} and \citet{mckee_temporal_2018} was used, de-trending using a running median  that is centred on a given epoch, $t_{\rm c}$, with a window in days denoted, $W$, 

\begin{equation}
    \tilde{x}(t_{\rm c}) = \text{med} \left[ x(t_{\rm i})~ | ~t_{\rm c} - \frac{W}{2} \leq t_{\rm i} \leq t_{\rm c} + \frac{W}{2} \right]\;,
\end{equation}
with the residuals then being $\Delta \text{DM}(t) = \text{DM}(t) - \widetilde{DM}(t)$ and $\Delta \tau = \tau (t) - \tilde \tau(t)$. The correlations for various windows are shown in \cref{fig:detrended_correlation}. The max point of correlation is for the thin model is $r = 0.48 \pm0.07$ at $W = 140$ d and $\rho = 0.52 \pm 0.07$ at $W = 190$ d, for the thick model $r = 0.49 \pm 0.08$ at $W = 125$ d and $\rho = 0.50 \pm 0.07$ at $W = 180$ d. These estimates are consistent with the 200 day de-trend timescale and correlation of 0.56 that \cite{mckee_temporal_2018} previously reported at higher frequency passbands. 

\begin{figure*}[t]
    \centering
    \includegraphics[width=0.8\linewidth]{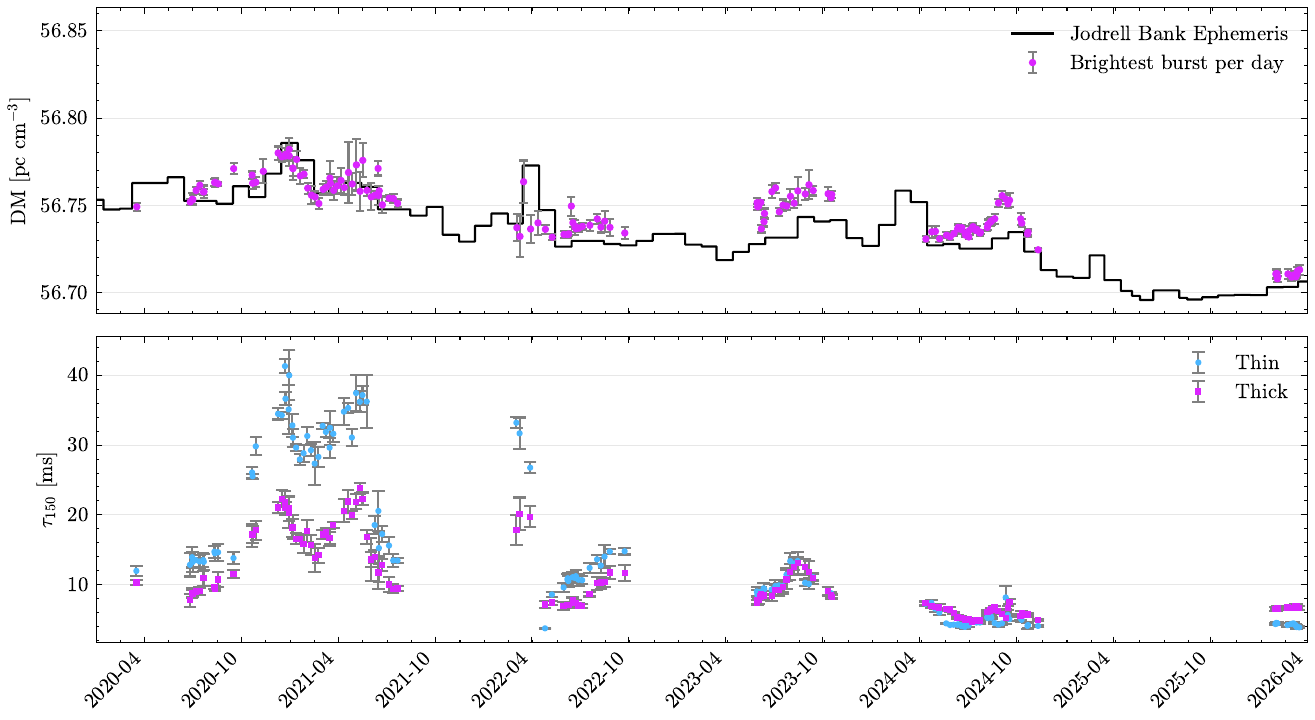}
    \caption{The \textit{top} panel shows the measured DM from 6 years of I-LOFAR data compared with ephemeris measured by Jodrell bank \citep{lyne_23_1993}. Whereas the \textit{(bottom)} shows the modelled scattering timescale for both the thick and modified thin screen models for each epoch during the observation campaign} 
    \label{fig:scatteringtDMtimeseries}
\end{figure*}

\subsection{Implications on the Crab Nebula}

As previously discussed, scattering timescales and dispersion measure are not independent observables. The time series itself regardless suggests that there are structures in the Crab Nebula itself that vary significantly on small timescales (i.e. 2020-10 to 2021-04, \cref{fig:scatteringtDMtimeseries}). 

\begin{figure}[htbp]
    \centering
    \includegraphics[width=0.96\linewidth]{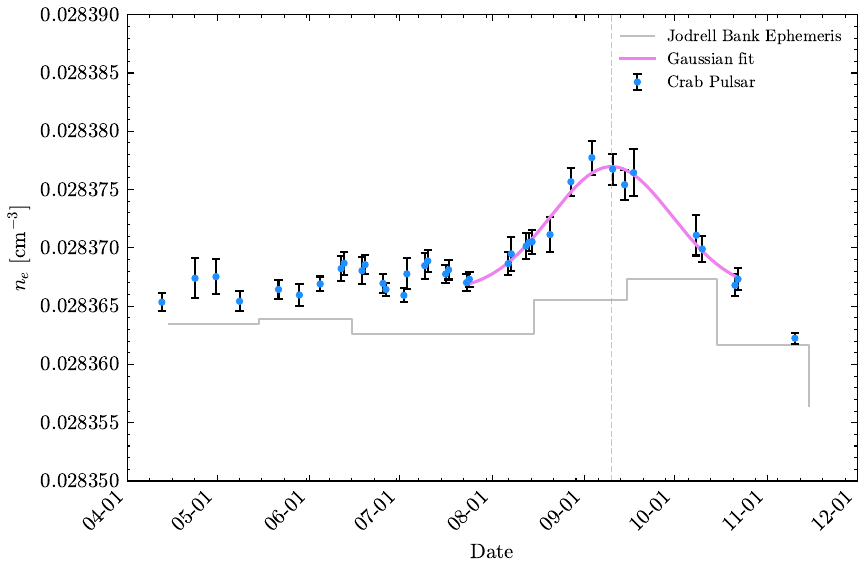}
    \caption{Electron density, $n_e$ as a function of time. With a Guassian fitted shown in pink and Jodrell bank Ephemeris shown in grey.}
    \label{fig:datevsNe}
\end{figure}

The DM and scattering time series (\cref{fig:scatteringtDMtimeseries}) are strongly correlated, with coefficients of $r = 0.71$ and $0.70$ for the thin (\cref{eq:thin_model_mod}) and thick (\cref{eq:thick_model}) screen models respectively. Part of this reflects a long-term decrease common to both DM and scattering over the past six years. De-trending with a running median of width 20--360 days, the correlation peaks at $r \approx 0.48$--$0.50$ ($\rho \approx 0.50$--$0.52$) for windows of $\sim$125--190 days, comparable to the $0.56 \pm 0.01$ found by \cite{mckee_temporal_2018} at 610~MHz. The cross-correlation of the de-trended series peaks at a lag consistent with zero, in agreement with the $-10 \pm 10$~day lag of \cite{mckee_temporal_2018} and, within the uncertainties, with the $\sim$30-day delay between scattering and DM peaks reported by \cite{kuzmin_deviation_2008} at 111~MHz. DM and scattering therefore vary together on timescales of months, consistent with discrete structures within the nebula varying the electron column density.

Individual DM events can be used to constrain the size of such structures. Between August and October 2024 the DM shows a distinct rise while the scattering remains reasonably low in turn minimally affecting DM measurements (\cref{fig:scatteringtDMtimeseries}). Fitting a Gaussian to the electron content over this interval, and assuming the excess arises from a single structure crossing the line of sight rather than from the background electron density, the crossing time is taken as the Gaussian FWHM, $\Delta t = 2.355\sigma$, with $\sigma = 20.22$~days, giving $\Delta t \approx 47.6$~days. The peak excess above the baseline corresponds to $\Delta\mathrm{DM} \approx 0.02~$\dm, similar to the $\sim0.03$\dm~variations reported by \cite{kuzmin_deviation_2008}.

If the variation is driven by the motion of the pulsar behind the nebular material, the relevant velocity is the pulsar's transverse velocity, $v_\perp \approx 120~\mathrm{km~s^{-1}}$ \citep{kaplan_precise_2008}. The transverse size of the structure is then $L \approx v_\perp \Delta t \approx 3.3$~AU ($\approx 1.6\times10^{-5}$~pc), comparable to the $\sim$6~AU structures inferred by \cite{mckee_temporal_2018} from the characteristic $\sim$70-day timescale of their DM variations. Even if the material itself moves at the $\sim$$10^{3}~\mathrm{km~s^{-1}}$ typical of the filament proper motions \citep{trimble_motions_1968}, $L \lesssim 30$~AU ($\approx 1.3\times10^{-4}$~pc). In either case the structure is two to three orders of magnitude smaller than the $\sim$0.03~pc dense cores and $\sim$0.08--0.2~pc diffuse envelopes of the Crab's filaments resolved by \cite{arias_crab_2025}; at the pulsar's velocity a filament core would take $\sim$250~years to cross the line of sight. Assuming a depth along the line of sight comparable to its transverse size, the local electron density of the structure is $n_e \approx \Delta\mathrm{DM}/L \sim 10^{2}$--$10^{3}~\mathrm{cm^{-3}}$.

Taken together, these constraints imply that the short timescale DM variations are produced neither by the passage of a whole filament nor by the optical wisps, whose lateral scales of $\sim$$1.6\times10^{-3}$~pc correspond to crossing times of $\sim$13~years \citep{mckee_temporal_2018}. It instead points to compact, AU-scale density substructure, well below the resolution of current imaging, such as fine structure within the Rayleigh--Taylor-shaped filamentary ejecta. Continued monitoring and analysis across many wavelengths events will allow the distribution of crossing times to place limits on the sizes and densities of these structures, and to test whether nearby compact plerionic structures contribute.

\subsection{Implications on the Solar Wind}

\begin{figure}
    \centering
    \includegraphics[width=0.9\linewidth]{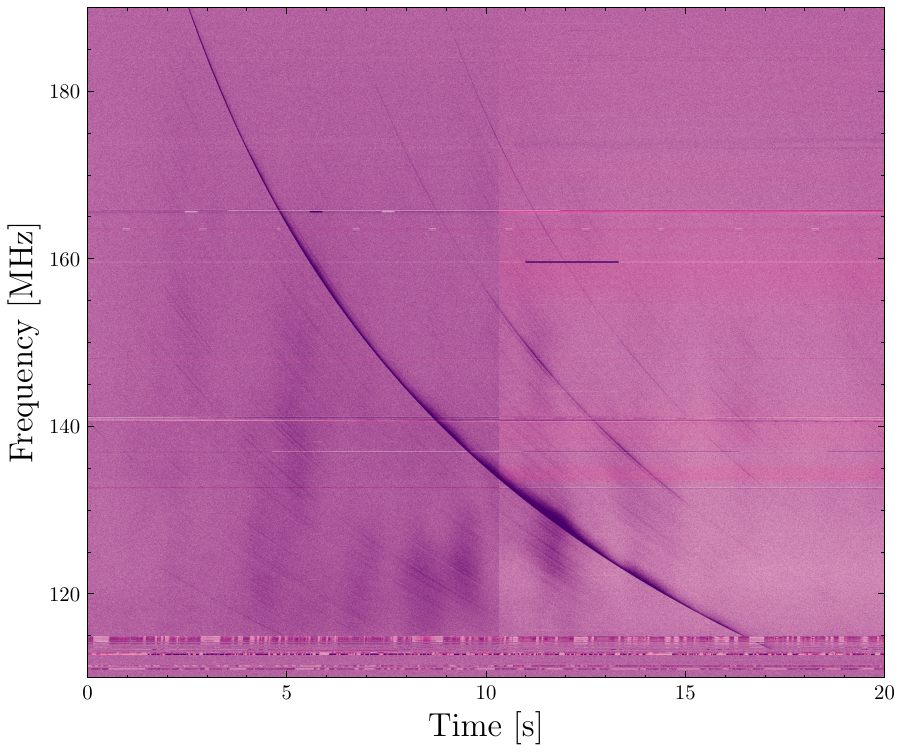}
    \caption{Dispersed waterfall of several GPs observed during the 30th of June 2020 with a solar separation of $14.9^\circ$ from beam bore sight with ionospheric scintillation present as initially shown in fig. 6 of \citet{murphy_first_2021}. The most prominent burst had an S/N of 1720.1 with a corrected DM of $56.754 \pm 0.003$ \dm. }
    \label{fig:GPthroughBurst}
\end{figure}

Probing the solar wind with GPs over long timescales proves to be more difficult, while some GPs can be clearly seen during radio solar bursts as shown in \cref{fig:GPthroughBurst}, the precision needed to probe the electron content from the Sun is on the order of $10^{-3}-10^{-4}$~\dm~\citep{tiburzi_impact_2021, susarla_exploring_2024}. The variations in DM caused by the Crab Nebula are too large to measure temporal variations in electron content caused by the solar wind. As expected the number of observed bursts decreases along with solar elongation (see table \ref{tab:bursts}).  

\section{Conclusion \& Future Work} \label{sec:conclusion_crab}

This work presents a six-year observing campaign of the Crab pulsar using I-LOFAR. The dataset comprises over a million single pulses and half a million giant pulses. The exceptional brightness of the Crab pulsar allowed for measurement of both dispersion measure and the pulse broadening due to scattering along the line of sight for over 5000 pulses above an S/N of 100.

A DM time series was created using both single pulse phase coherent DM optimisation and timing through folding. It was found that apparent DM offsets can arise from frequency-dependent pulse broadening rather than genuine changes in the DM, making it difficult to determine any variation in the local nebula on short timescales in comparison to other higher frequency Crab observations. Although lower frequencies provide enhanced sensitivity to small DM offsets, the rapidly increasing scattering broadens the pulse profile sufficiently to limit the achievable DM precision from individual giant pulses.


A way to account for this could be the use of cyclic spectroscopy \citep{demorest_cyclic_2011,dolch_deconvolving_2021,turner_constraining_2026}. Cyclic spectroscopy is a coherent folding technique that recovers phase-preserving, high-resolution frequency information by exploiting pulsar periodicity. By treating the ISM as a linear transfer function that convolves the intrinsic pulsar voltage signal, the technique uses the pulsar's periodic coherence to estimate and, where feasible, deconvolve the ISM impulse response. This allows the intrinsic pulse phase structure to be recovered and scattering-induced broadening to be corrected. Such a correction would make it easier to discern the true phase of individual pulses, as the phase boundary is currently too `fuzzy' to allow a sharp separation between pulse populations. This method was not implemented in this study, as it is applied directly to raw voltages, whereas only 32-bit filterbanks were archived. Furthermore, its application would be complicated by temporal variations in the local nebula itself. Nevertheless, correcting even a fraction of the scattering may sufficiently improve DM precision to allow further modelling of the local nebula's electron content. At the time of writing, there is a cyclic spectroscopy backend being commissioned at the Robert C. Byrd Green Bank Telescope \citep{Lynch2026GBTMemo324} the results of these developments will indicate inclusion on future signal processing chain on I-LOFAR. 

The measured DM and scattering variations further suggest that the Crab Nebula is highly dynamic on observable timescales, with turbulent or filamentary structures contributing significantly to the observed propagation effects. It was also determined that the local variations in the nebula aren't well enough constrained or modelled to be able to `untangle' DM changes which limits the ability to probe the solar wind and obtain an accurate ephemeris for timing at low frequencies.

Analysis of the giant-pulse population indicated that the measured S/N, flux density and fluence distributions follow power-law behaviour consistent with previous studies of the Crab pulsar. Although this study noted an excess in each of the distributions in addition to not completely following a power law. This is primarily attributed to the difficulty in the separation of the main and inter pulse populations due the effects of the varied scattering. As noted throughout the study scattering timescales vary from pulse to pulse and also over the entire observation campaign. 

Combining the long-term DM and scattering measurements also enabled a discussion on the potential constraints that low frequency observations could place on the local Crab environment. The duration and magnitude of a specific DM enhancement in late 2024/early 2025 are consistent with a diffuse filamentary or Rayleigh-Taylor ejecta structure crossing the line of sight. Although the exact electron-density distribution cannot yet be uniquely determined, these measurements demonstrate the potential of long-term low-frequency giant-pulse monitoring to probe the evolving plasma environment surrounding the Crab pulsar.

Overall, this work demonstrates that the study of giant pulses from the Crab pulsar at low frequencies as one of the most powerful probes of the intervening medium along the line of sight. Continued long-term monitoring and simultaneous observations at higher frequencies will provide a more complete picture of the evolving electron content of the Crab Nebula and improve our understanding of pulse propagation through complex plasma environments.












\section*{Acknowledgements}
OAJ acknowledges the support of the Trinity College Dublin, School of Physics and Breakthrough Listen which is managed by the Breakthrough Prize Foundation. LV acknowledges the support of a Lindsay Scholarship, funded by Armagh Observatory \& Planetarium, Dublin Institute for Advanced Studies and Trinity College Dublin. The Rosse Observatory is operated by Trinity College Dublin; we acknowledge support from a Higher Education Research Equipment Grant (PI: Keane). I-LOFAR infrastructure has benefited from funding from Science Foundation Ireland, a predecessor of Taighde Éireann --- Research Ireland. The authors would like to thank Oleg Smirnov and Kate Maguire for their comments and insights.

\medskip\noindent\textit{Author contributions.}
The work presented here uses data collected using the Irish LOFAR telescope ($\sim \tobs$ hours) over the course of 6 years. The work resulted in the development of much software, as acknowledged in the main text. It also benefited from, and built upon, a large pre-existing software base developed by many authors over many years, in particular software developers in the pulsar science community and at ASTRON.

Contributions by individuals specific to the work presented in this paper, as opposed to the ongoing effort to operate the telescopes, are listed below in the CRedIT\footnote{\url{https://credit.niso.org/}} format.

\medskip\noindent\textit{CRediT statement:}\\
OAJ: Conceptualization, Methodology, Software, Validation, Formal Analysis, Investigation, Data Curation, Writing --- Original Draft, Writing --- Review \& Editing, Visualization. \\
DJMcK: Conceptualisation, Software, Resources, Writing --- Review \& Editing. \\
SCS: Software, Investigation, Data Curation, Writing — Review \& Editing \\
EFK: Conceptualization, Methodology, Investigation, Resources, Writing — Original Draft, Writing — Review \& Editing, Supervision, Project Administration, Funding Acquisition. \\
LV: Software, Resources \\
JMcC: Software, Resources, Project Administration \\ 
AG: Conceptualization, Resources, Review \& Editing 


\bibliography{crab,software,hardware}

@article{jessner_giant_2010,
	title = {Giant pulses with nanosecond time resolution detected from the {Crab} pulsar at 8.5 and 15.1 {GHz}},
	volume = {524},
	copyright = {© ESO, 2010},
	issn = {0004-6361, 1432-0746},
	url = {https://www.aanda.org/articles/aa/abs/2010/16/aa14806-10/aa14806-10.html},
	doi = {10.1051/0004-6361/201014806},
	language = {en},
	urldate = {2026-01-20},
	journal = {Astronomy \& Astrophysics},
	author = {Jessner, A. and Popov, M. V. and Kondratiev, V. I. and Kovalev, Y. Y. and Graham, D. and Zensus, A. and Soglasnov, V. A. and Bilous, A. V. and Moshkina, O. A.},
	month = dec,
	year = {2010},
	note = {Publisher: EDP Sciences},
	pages = {A60},
}

@article{thulasiram_narrow-band_2021,
	title = {Narrow-band giant pulses from the {Crab} pulsar},
	volume = {508},
	issn = {0035-8711},
	url = {https://doi.org/10.1093/mnras/stab2692},
	doi = {10.1093/mnras/stab2692},
	number = {2},
	urldate = {2026-01-20},
	journal = {Monthly Notices of the Royal Astronomical Society},
	author = {Thulasiram, Parasar and Lin, Hsiu-Hsien},
	month = dec,
	year = {2021},
	pages = {1947--1953},
}

@article{bij_kinematics_2021,
	title = {Kinematics of {Crab} {Giant} {Pulses}},
	volume = {920},
	issn = {0004-637X},
	url = {https://ui.adsabs.harvard.edu/abs/2021ApJ...920...38B},
	doi = {10.3847/1538-4357/ac1589},
	urldate = {2026-01-20},
	journal = {The Astrophysical Journal},
	author = {Bij, Akanksha and Lin, Hsiu-Hsien and Li, Dongzi and van Kerkwijk, Marten H. and Pen, Ue-Li and Lu, Wenbin and Main, Robert and Peterson, Jeffrey B. and Quine, Brendan and Vanderlinde, Keith},
	month = oct,
	year = {2021},
	note = {Publisher: IOP
ADS Bibcode: 2021ApJ...920...38B},
	pages = {38},
}

@article{karuppusamy_giant_2010,
	title = {Giant pulses from the {Crab} pulsar. {A} wide-band study},
	volume = {515},
	issn = {0004-6361},
	url = {https://ui.adsabs.harvard.edu/abs/2010A&A...515A..36K},
	doi = {10.1051/0004-6361/200913729},
	urldate = {2026-01-20},
	journal = {Astronomy and Astrophysics},
	author = {Karuppusamy, R. and Stappers, B. W. and van Straten, W.},
	month = jun,
	year = {2010},
	note = {Publisher: EDP
ADS Bibcode: 2010A\&A...515A..36K},
	pages = {A36},
}

@article{sokolowski_100_2025,
	title = {100 000 {Crab} giant pulses at 215 {MHz} detected with an {SKA}-{Low} prototype station},
	volume = {42},
	issn = {1323-3580},
	url = {https://ui.adsabs.harvard.edu/abs/2025PASA...42..145S},
	doi = {10.1017/pasa.2025.10108},
	urldate = {2026-01-20},
	journal = {Publications of the Astronomical Society of Australia},
	author = {Sokolowski, Marcin and Kumar, Pratik and Dhavali, Shrada and Meyers, Bradley W. and Bhat, Ramesh and Bera, Apurba and McSweeney, Samuel},
	month = oct,
	year = {2025},
	note = {ADS Bibcode: 2025PASA...42..145S},
	pages = {e145},
}

@article{cordes_brightest_2004,
	title = {The {Brightest} {Pulses} in the {Universe}: {Multifrequency} {Observations} of the {Crab} {Pulsar}'s {Giant} {Pulses}},
	volume = {612},
	issn = {0004-637X, 1538-4357},
	shorttitle = {The {Brightest} {Pulses} in the {Universe}},
	url = {http://arxiv.org/abs/astro-ph/0304495},
	doi = {10.1086/422495},
	number = {1},
	urldate = {2026-01-20},
	journal = {The Astrophysical Journal},
	author = {Cordes, J. M. and Bhat, N. D. R. and Hankins, T. H. and McLaughlin, M. A. and Kern, J.},
	month = sep,
	year = {2004},
	note = {arXiv:astro-ph/0304495},
	pages = {375--388},
}

@article{lyubarsky_structure_2002,
	title = {On the structure of the inner {Crab} {Nebula}},
	volume = {329},
	issn = {0035-8711},
	url = {https://doi.org/10.1046/j.1365-8711.2002.05151.x},
	doi = {10.1046/j.1365-8711.2002.05151.x},
	number = {2},
	urldate = {2026-01-28},
	journal = {Monthly Notices of the Royal Astronomical Society},
	author = {Lyubarsky, Y. E.},
	month = jan,
	year = {2002},
	pages = {L34--L36},
}

@article{gaensler_evolution_2006,
	title = {The {Evolution} and {Structure} of {Pulsar} {Wind} {Nebulae}},
	volume = {44},
	issn = {0066-4146},
	url = {https://ui.adsabs.harvard.edu/abs/2006ARA&A..44...17G},
	doi = {10.1146/annurev.astro.44.051905.092528},
	urldate = {2026-01-28},
	journal = {Annual Review of Astronomy and Astrophysics},
	author = {Gaensler, Bryan M. and Slane, Patrick O.},
	month = sep,
	year = {2006},
	note = {ADS Bibcode: 2006ARA\&A..44...17G},
	pages = {17--47},
}

@article{hester_crab_2008,
	title = {The {Crab} {Nebula} : an astrophysical chimera.},
	volume = {46},
	issn = {0066-4146},
	shorttitle = {The {Crab} {Nebula}},
	url = {https://ui.adsabs.harvard.edu/abs/2008ARA&A..46..127H},
	doi = {10.1146/annurev.astro.45.051806.110608},
	urldate = {2026-01-28},
	journal = {Annual Review of Astronomy and Astrophysics},
	author = {Hester, J. J.},
	month = sep,
	year = {2008},
	note = {ADS Bibcode: 2008ARA\&A..46..127H},
	pages = {127--155},
}

@article{kuzmin_deviation_2008,
	title = {Deviation of the arrival-time delay of pulses for the {Crab} pulsar from a quadratic frequency region},
	volume = {52},
	issn = {0004-62991063-7729},
	url = {https://ui.adsabs.harvard.edu/abs/2008ARep...52..910K},
	doi = {10.1134/S106377290811005X},
	urldate = {2026-01-28},
	journal = {Astronomy Reports},
	author = {Kuz'min, A. D. and Losovskii, B. Ya. and Logvinenko, S. V. and Litvinov, I. I.},
	month = nov,
	year = {2008},
	note = {Publisher: Springer
ADS Bibcode: 2008ARep...52..910K},
	pages = {910--916},
}

@article{ATNFCat_2005,
	title = {The {Australia} {Telescope} {National} {Facility} {Pulsar} {Catalogue}},
	volume = {129},
	issn = {0004-6256},
	url = {https://ui.adsabs.harvard.edu/abs/2005AJ....129.1993M},
	doi = {10.1086/428488},
	urldate = {2025-05-19},
	journal = {The Astronomical Journal},
	author = {Manchester, R. N. and Hobbs, G. B. and Teoh, A. and Hobbs, M.},
	month = apr,
	year = {2005},
	note = {Publisher: IOP
ADS Bibcode: 2005AJ....129.1993M},
	pages = {1993--2006},
}

@article{isaacman_crab_1977,
	title = {The {Crab} nebula pulsar: variability of dispersion and scattering.},
	volume = {214},
	issn = {0004-637X},
	shorttitle = {The {Crab} nebula pulsar},
	url = {https://ui.adsabs.harvard.edu/abs/1977ApJ...214..214I},
	doi = {10.1086/155245},
	urldate = {2026-01-29},
	journal = {The Astrophysical Journal},
	author = {Isaacman, R. and Rankin, J. M.},
	month = may,
	year = {1977},
	note = {Publisher: IOP
ADS Bibcode: 1977ApJ...214..214I},
	pages = {214--232},
}

@article{iraci_pulsar_2024,
	title = {Pulsar timing methods for evaluating dispersion measure time series},
	volume = {692},
	issn = {0004-6361},
	url = {https://ui.adsabs.harvard.edu/abs/2024A&A...692A.170I},
	doi = {10.1051/0004-6361/202450740},
	urldate = {2026-01-30},
	journal = {Astronomy and Astrophysics},
	publisher = {EDP},
	author = {Iraci, F. and Chalumeau, A. and Tiburzi, C. and Verbiest, J. P. W. and Possenti, A. and Shaifullah, G. M. and Susarla, S. C. and Krishnakumar, M. A. and Lam, M. T. and Cromartie, H. T. and Kerr, M. and Grießmeier, Jean-Mathias},
	month = dec,
	year = {2024},
	note = {ADS Bibcode: 2024A\&A...692A.170I},
	pages = {A170},
}

@ARTICLE{sr68,
       author = {{Staelin}, David H. and {Reifenstein}, III, Edward C.},
        title = "{Pulsating Radio Sources near the Crab Nebula}",
      journal = {Science},
         year = 1968,
        month = dec,
       volume = {162},
       number = {3861},
        pages = {1481-1483},
          doi = {10.1126/science.162.3861.1481},
       adsurl = {https://ui.adsabs.harvard.edu/abs/1968Sci...162.1481S}
}

@ARTICLE{hje15,
       author = {{Hankins}, T.~H. and {Jones}, G. and {Eilek}, J.~A.},
        title = "{The Crab Pulsar at Centimeter Wavelengths. I. Ensemble Characteristics}",
      journal = {\apj},
         year = 2015,
        month = apr,
       volume = {802},
       number = {2},
          eid = {130},
        pages = {130},
          doi = {10.1088/0004-637X/802/2/130},
archivePrefix = {arXiv},
       eprint = {1502.00677},
 primaryClass = {astro-ph.HE},
       adsurl = {https://ui.adsabs.harvard.edu/abs/2015ApJ...802..130H}
}

@article{seymour_dm_phase_2019,
    title = {{DM}\_phase: {Algorithm} for correcting dispersion of radio signals},
    shorttitle = {{DM}\_phase},
    url = {https://ui.adsabs.harvard.edu/abs/2019ascl.soft10004S},
    urldate = {2026-02-23},
    journal = {Astrophysics Source Code Library},
    author = {Seymour, Andrew and Michilli, Daniele and Pleunis, Ziggy},
    month = oct,
    year = {2019},
    note = {ADS Bibcode: 2019ascl.soft10004S},
    pages = {ascl:1910.004},
}

@article{lyne_23_1993,
    title = {23 years of {Crab} pulsar rotational history.},
    volume = {265},
    issn = {0035-8711},
    url = {https://ui.adsabs.harvard.edu/abs/1993MNRAS.265.1003L},
    doi = {10.1093/mnras/265.4.1003},
    urldate = {2026-03-02},
    journal = {Monthly Notices of the Royal Astronomical Society},
    publisher = {OUP},
    author = {Lyne, A. G. and Pritchard, R. S. and Graham Smith, F.},
    month = dec,
    year = {1993},
    note = {ADS Bibcode: 1993MNRAS.265.1003L},
    pages = {1003--1012},
}

@article{hotan_psrchive_2004,
    title = {{PSRCHIVE} and {PSRFITS}: {An} {Open} {Approach} to {Radio} {Pulsar} {Data} {Storage} and {Analysis}},
    volume = {21},
    issn = {1323-3580},
    shorttitle = {{PSRCHIVE} and {PSRFITS}},
    url = {https://ui.adsabs.harvard.edu/abs/2004PASA...21..302H},
    doi = {10.1071/AS04022},
    urldate = {2026-03-02},
    journal = {Publications of the Astronomical Society of Australia},
    author = {Hotan, A. W. and van Straten, W. and Manchester, R. N.},
    month = jan,
    year = {2004},
    note = {ADS Bibcode: 2004PASA...21..302H},
    pages = {302--309},
}

@article{bhat_detection_2007,
    title = {Detection of {Crab} {Giant} {Pulses} {Using} the {Mileura} {Widefield} {Array} {Low} {Frequency} {Demonstrator} {Field} {Prototype} {System}},
    volume = {665},
    issn = {0004-637X},
    url = {https://ui.adsabs.harvard.edu/abs/2007ApJ...665..618B},
    doi = {10.1086/519444},
    urldate = {2026-03-04},
    journal = {The Astrophysical Journal},
    publisher = {IOP},
    author = {Bhat, N. D. Ramesh and Wayth, Randall B. and Knight, Haydon S. and Bowman, Judd D. and Oberoi, Divya and Barnes, David G. and Briggs, Frank H. and Cappallo, Roger J. and Herne, David and Kocz, Jonathon and Lonsdale, Colin J. and Lynch, Mervyn J. and Stansby, Bruce and Stevens, Jamie and Torr, Glen and Webster, Rachel L. and Wyithe, J. Stuart B.},
    month = aug,
    year = {2007},
    note = {ADS Bibcode: 2007ApJ...665..618B},
    pages = {618--627},
}

@article{bhat_clean-based_2003,
    title = {A {CLEAN}-based {Method} for {Deconvolving} {Interstellar} {Pulse} {Broadening} from {Radio} {Pulses}},
    volume = {584},
    issn = {0004-637X},
    url = {https://ui.adsabs.harvard.edu/abs/2003ApJ...584..782B},
    doi = {10.1086/345775},
    urldate = {2026-03-04},
    journal = {The Astrophysical Journal},
    publisher = {IOP},
    author = {Bhat, N. D. R. and Cordes, J. M. and Chatterjee, S.},
    month = feb,
    year = {2003},
    note = {ADS Bibcode: 2003ApJ...584..782B},
    pages = {782--790},
}

@article{williamson_pulse_1972,
    title = {Pulse {Broadening} due to {Multiple} {Scattering} in the {Interstellar} {Medium}},
    volume = {157},
    issn = {0035-8711},
    url = {https://doi.org/10.1093/mnras/157.1.55},
    doi = {10.1093/mnras/157.1.55},
    number = {1},
    urldate = {2026-03-04},
    journal = {Monthly Notices of the Royal Astronomical Society},
    author = {Williamson, Ian P.},
    month = apr,
    year = {1972},
    pages = {55--71},
}

@article{susarla_long-term_2025,
    title = {Long-term timing results of ecliptic pulsars observed with {I}-{LOFAR}},
    volume = {698},
    issn = {0004-6361},
    url = {https://ui.adsabs.harvard.edu/abs/2025A&A...698A.248S},
    doi = {10.1051/0004-6361/202554226},
    urldate = {2026-01-29},
    journal = {Astronomy and Astrophysics},
    publisher = {EDP},
    author = {Susarla, S. C. and Johnson, O. A. and McKenna, D. J. and Keane, E. F. and McCauley, P. J. and Verbiest, J. P. W. and Tiburzi, C. and Golden, A.},
    month = jun,
    year = {2025},
    note = {ADS Bibcode: 2025A\&A...698A.248S},
    pages = {A248},
}

@misc{johnson_radio_2025,
    title = {Radio {Observations} of a {Candidate} {Redback} {Millisecond} {Pulsar}: {1FGL} {J0523}.5-2529},
    copyright = {Creative Commons Attribution-NonCommercial-NoDerivatives 4.0 International License},
    shorttitle = {Radio {Observations} of a {Candidate} {Redback} {Millisecond} {Pulsar}},
    url = {http://arxiv.org/abs/2508.15435},
    doi = {10.48550/arXiv.2508.15435},
    urldate = {2025-11-18},
    publisher = {arXiv},
    author = {Johnson, O. A. and Keane, E. F. and McKenna, D. J. and Qiu, H. and Swihart, S. J. and Strader, J. and McLaughlin, M.},
    month = nov,
    year = {2025},
    note = {arXiv:2508.15435 [astro-ph]},
}

@article{mckenna_census_2024,
    title = {A census of rotating radio transients at 150 {MHz} with the {Irish} {LOFAR} station},
    volume = {527},
    issn = {0035-8711},
    url = {https://doi.org/10.1093/mnras/stad2900},
    doi = {10.1093/mnras/stad2900},
    number = {3},
    urldate = {2026-05-11},
    journal = {Monthly Notices of the Royal Astronomical Society},
    author = {McKenna, D J and Keane, E F and Gallagher, P T and McCauley, J},
    month = jan,
    year = {2024},
    pages = {4397--4419},
}

@article{demorest_cyclic_2011,
    title = {Cyclic spectral analysis of radio pulsars},
    volume = {416},
    issn = {0035-8711},
    url = {https://doi.org/10.1111/j.1365-2966.2011.19230.x},
    doi = {10.1111/j.1365-2966.2011.19230.x},
    number = {4},
    urldate = {2026-05-13},
    journal = {Monthly Notices of the Royal Astronomical Society},
    author = {Demorest, P. B.},
    month = oct,
    year = {2011},
    pages = {2821--2826},
}

@article{dolch_deconvolving_2021,
    title = {Deconvolving {Pulsar} {Signals} with {Cyclic} {Spectroscopy}: {A} {Systematic} {Evaluation}},
    volume = {913},
    issn = {0004-637X},
    shorttitle = {Deconvolving {Pulsar} {Signals} with {Cyclic} {Spectroscopy}},
    url = {https://doi.org/10.3847/1538-4357/abf48b},
    doi = {10.3847/1538-4357/abf48b},
    language = {en},
    number = {2},
    urldate = {2026-05-13},
    journal = {The Astrophysical Journal},
    publisher = {The American Astronomical Society},
    author = {Dolch, Timothy and Stinebring, Dan R. and Jones, Glenn and Zhu, Hengrui and Lynch, Ryan S. and Cohen, Tyler and Demorest, Paul B. and Lam, Michael T. and Levin, Lina and McLaughlin, Maura A. and Palliyaguru, Nipuni T.},
    month = may,
    year = {2021},
    pages = {98},
}

@article{susarla_exploring_2024,
    title = {Exploring the time variability of the solar wind using {LOFAR} pulsar data},
    volume = {692},
    issn = {0004-6361},
    url = {https://ui.adsabs.harvard.edu/abs/2024A&A...692A..18S},
    doi = {10.1051/0004-6361/202450680},
    urldate = {2026-01-29},
    journal = {Astronomy and Astrophysics},
    publisher = {EDP},
    author = {Susarla, S. C. and Chalumeau, A. and Tiburzi, C. and Keane, E. F. and Verbiest, J. P. W. and Hazboun, J. S. and Krishnakumar, M. A. and Iraci, F. and Shaifullah, G. M. and Golden, A. and Bak Nielsen, A.-S. and Donner, J. and Grießmeier, J.-M. and Keith, M. J. and Osłowski, S. and Porayko, N. K. and Serylak, M. and Anderson, J. M. and Brüggen, M. and Ciardi, B. and Dettmar, R.-J. and Hoeft, M. and Künsemöller, J. and Schwarz, D. and Vocks, C.},
    month = dec,
    year = {2024},
    note = {ADS Bibcode: 2024A\&A...692A..18S},
    pages = {A18},
}

@article{tiburzi_impact_2021,
    title = {The impact of solar wind variability on pulsar timing},
    volume = {647},
    issn = {0004-6361},
    url = {https://ui.adsabs.harvard.edu/abs/2021A&A...647A..84T},
    doi = {10.1051/0004-6361/202039846},
    urldate = {2026-05-15},
    journal = {Astronomy and Astrophysics},
    publisher = {EDP},
    author = {Tiburzi, C. and Shaifullah, G. M. and Bassa, C. G. and Zucca, P. and Verbiest, J. P. W. and Porayko, N. K. and van der Wateren, E. and Fallows, R. A. and Main, R. A. and Janssen, G. H. and Anderson, J. M. and Bak Nielsen, A.-S. and Donner, J. Y. and Keane, E. F. and Künsemöller, J. and Osłowski, S. and Grießmeier, J.-M. and Serylak, M. and Brüggen, M. and Ciardi, B. and Dettmar, R.-J. and Hoeft, M. and Kramer, M. and Mann, G. and Vocks, C.},
    month = mar,
    year = {2021},
    note = {ADS Bibcode: 2021A\&A...647A..84T},
    pages = {A84},
}

@article{hobbs_tempo2_2006,
    title = {{TEMPO2}, a new pulsar-timing package - {I}. {An} overview},
    volume = {369},
    issn = {0035-8711},
    url = {https://ui.adsabs.harvard.edu/abs/2006MNRAS.369..655H},
    doi = {10.1111/j.1365-2966.2006.10302.x},
    urldate = {2026-05-18},
    journal = {Monthly Notices of the Royal Astronomical Society},
    publisher = {OUP},
    author = {Hobbs, G. B. and Edwards, R. T. and Manchester, R. N.},
    month = jun,
    year = {2006},
    note = {ADS Bibcode: 2006MNRAS.369..655H},
    pages = {655--672},
}

@article{lorimer_sigproc_2011,
    title = {{SIGPROC}: {Pulsar} {Signal} {Processing} {Programs}},
    shorttitle = {{SIGPROC}},
    url = {https://ui.adsabs.harvard.edu/abs/2011ascl.soft07016L},
    urldate = {2026-05-19},
    journal = {Astrophysics Source Code Library},
    author = {Lorimer, D. R.},
    month = jul,
    year = {2011},
    note = {ADS Bibcode: 2011ascl.soft07016L},
    pages = {ascl:1107.016},
}

@article{mckee_temporal_2018,
    title = {Temporal variations in scattering and dispersion measure in the {Crab} {Pulsar} and their effect on timing precision},
    volume = {479},
    issn = {0035-8711},
    url = {https://doi.org/10.1093/mnras/sty1727},
    doi = {10.1093/mnras/sty1727},
    number = {3},
    urldate = {2026-05-22},
    journal = {Monthly Notices of the Royal Astronomical Society},
    author = {McKee, J W and Lyne, A G and Stappers, B W and Bassa, C G and Jordan, C A},
    month = sep,
    year = {2018},
    pages = {4216--4224},
}

@article{akaike_new_1974,
  title = {A {New} {Look} at the {Statistical} {Model} {Identification}},
  volume = {19},
  issn = {0018-9286},
  url = {https://ui.adsabs.harvard.edu/abs/1974ITAC...19..716A},
  doi = {10.1109/TAC.1974.1100705},
  urldate = {2026-10-27},
  journal = {IEEE Transactions on Automatic Control},
  publisher = {IEEE},
  author = {Akaike, H.},
  month = jan,
  year = {1974},
  note = {ADS Bibcode: 1974ITAC...19..716A},
  pages = {716--723},
}

@article{arias_crab_2025,
    title = {The {Crab} {Nebula} at subarcsecond resolution with the {International} {LOFAR} {Telescope}},
    volume = {699},
    issn = {0004-6361, 1432-0746},
    url = {http://arxiv.org/abs/2506.19460},
    doi = {10.1051/0004-6361/202555333},
    urldate = {2026-07-30},
    journal = {Astronomy \& Astrophysics},
    author = {Arias, M. and Timmerman, R. and Sweijen, F. and Weeren, R. J. van and Bassa, C. G.},
    month = jul,
    year = {2025},
    note = {arXiv:2506.19460 [astro-ph.HE]},
    pages = {A319},
}

@article{clauset_power-law_2009,
  title = {Power-{Law} {Distributions} in {Empirical} {Data}},
  volume = {51},
  issn = {0036-1445},
  url = {https://ui.adsabs.harvard.edu/abs/2009SIAMR..51..661C},
  doi = {10.1137/070710111},
  urldate = {2026-10-29},
  journal = {SIAM Review},
  author = {Clauset, Aaron and Shalizi, Cosma Rohilla and Newman, M. E. J.},
  month = jan,
  year = {2009},
  note = {ADS Bibcode: 2009SIAMR..51..661C},
  pages = {661--703},
}

@PHDTHESIS{david_thesis,
	address = {Dublin},
	title = {Characterising {Pulsars} with {Low}-{Frequency} {Software} {Telescopes}},
	school = {DIAS/Trinity College Dublin},
	author = {McKenna, David J.},
	month = sep,
	year = {2024},
}

@article{driessen_scattering_2019,
  title = {Scattering features and variability of the {Crab} pulsar},
  volume = {483},
  issn = {0035-8711},
  url = {https://doi.org/10.1093/mnras/sty3192},
  doi = {10.1093/mnras/sty3192},
  number = {1},
  urldate = {2026-10-27},
  journal = {Monthly Notices of the Royal Astronomical Society},
  author = {Driessen, L N and Janssen, G H and Bassa, C G and Stappers, B W and Stinebring, D R},
  month = feb,
  year = {2019},
  pages = {1224--1232},
}

@article{kaplan_precise_2008,
  title = {A {Precise} {Proper} {Motion} for the {Crab} {Pulsar}, and the {Difficulty} of {Testing} {Spin}-{Kick} {Alignment} for {Young} {Neutron} {Stars}},
  volume = {677},
  issn = {0004-637X},
  url = {https://ui.adsabs.harvard.edu/abs/2008ApJ...677.1201K},
  doi = {10.1086/529026},
  urldate = {2026-11-18},
  journal = {The Astrophysical Journal},
  publisher = {IOP},
  author = {Kaplan, D. L. and Chatterjee, S. and Gaensler, B. M. and Anderson, J.},
  month = apr,
  year = {2008},
  note = {ADS Bibcode: 2008ApJ...677.1201K},
  pages = {1201--1215},
}

@article{karuppusamy_crab_2012,
    title = {Crab giant pulses at low frequencies},
    volume = {538},
    issn = {0004-6361},
    url = {https://ui.adsabs.harvard.edu/abs/2012A&A...538A...7K},
    doi = {10.1051/0004-6361/201117667},
    urldate = {2026-02-26},
    journal = {Astronomy and Astrophysics},
    publisher = {EDP},
    author = {Karuppusamy, R. and Stappers, B. W. and Lee, K. J.},
    month = feb,
    year = {2012},
    note = {ADS Bibcode: 2012A\&A...538A...7K},
    pages = {A7},
}

@article{lam_systematic_2016,
  title = {{SYSTEMATIC} {AND} {STOCHASTIC} {VARIATIONS} {IN} {PULSAR} {DISPERSION} {MEASURES}},
  volume = {821},
  issn = {0004-637X},
  url = {https://doi.org/10.3847/0004-637X/821/1/66},
  doi = {10.3847/0004-637X/821/1/66},
  language = {en},
  number = {1},
  urldate = {2026-11-17},
  journal = {The Astrophysical Journal},
  publisher = {The American Astronomical Society},
  author = {Lam, M. T. and Cordes, J. M. and Chatterjee, S. and Jones, M. L. and McLaughlin, M. A. and Armstrong, J. W.},
  month = apr,
  year = {2016},
  pages = {66},
}

@article{petrova_physics_2008,
    title = {Physics of {Interpulse} {Emission} in {Radio} {Pulsars}},
    volume = {673},
    issn = {0004-637X},
    url = {https://ui.adsabs.harvard.edu/abs/2008ApJ...673..400P},
    doi = {10.1086/523693},
    urldate = {2026-07-30},
    journal = {The Astrophysical Journal},
    publisher = {IOP},
    author = {Petrova, S. A.},
    month = jan,
    year = {2008},
    note = {ADS Bibcode: 2008ApJ...673..400P},
    pages = {400--410},
}

@article{trimble_motions_1968,
    title = {Motions and {Structure} of the {Filamentary} {Envelope} of the {Crab} {Nebula}},
    volume = {73},
    issn = {0004-6256},
    url = {https://ui.adsabs.harvard.edu/abs/1968AJ.....73..535T},
    doi = {10.1086/110658},
    urldate = {2026-07-30},
    journal = {The Astronomical Journal},
    publisher = {IOP},
    author = {Trimble, Virginia},
    month = sep,
    year = {1968},
    note = {ADS Bibcode: 1968AJ.....73..535T},
    pages = {535},
}

@misc{turner_constraining_2026,
    title = {Constraining {Scattering} {Medium} {Geometry} with {Cyclic} {Spectroscopy}},
    url = {http://arxiv.org/abs/2605.26229},
    doi = {10.48550/arXiv.2605.26229},
    urldate = {2026-07-30},
    publisher = {arXiv},
    author = {Turner, Jacob E.},
    month = may,
    year = {2026},
    note = {arXiv:2605.26229 [astro-ph.HE]},
}

@misc{vincetti_characterising_2026,
  title = {Characterising the response of an {International} {LOFAR} {Station}},
  url = {http://arxiv.org/abs/2607.19585},
  doi = {10.48550/arXiv.2607.19585},
  urldate = {2026-07-23},
  publisher = {arXiv},
  author = {Vincetti, Letizia and Susarla, Sai C. and Johnson, Owen A. and Keane, Evan F. and McKenna, David J. and Carozzi, Tobia D. and Ramsay, Gavin and Creaner, Oisin and Gallagher, Peter T. and McCauley, Joe},
  month = jul,
  year = {2026},
  note = {arXiv:2607.19585 [astro-ph.IM]},
}

@article{ransom_presto_2011,
  title = {{PRESTO}: {PulsaR} {Exploration} and {Search} {TOolkit}},
  shorttitle = {{PRESTO}},
  url = {https://ui.adsabs.harvard.edu/abs/2011ascl.soft07017R},
  urldate = {2025-05-22},
  journal = {Astrophysics Source Code Library},
  author = {Ransom, Scott},
  month = jul,
  year = {2011},
  note = {ADS Bibcode: 2011ascl.soft07017R},
  pages = {ascl:1107.017},
}

@article{bietenholz_new_2015,
  title = {New expansion rate measurements of the {Crab} nebula in radio and optical},
  volume = {454},
  issn = {0035-8711},
  url = {https://doi.org/10.1093/mnras/stv2112},
  doi = {10.1093/mnras/stv2112},
  number = {3},
  urldate = {2026-11-20},
  journal = {Monthly Notices of the Royal Astronomical Society},
  author = {Bietenholz, M. F. and Nugent, R. L.},
  month = dec,
  year = {2015},
  pages = {2416--2422},
}

@ARTICLE{ljg+15,
       author = {{Lyne}, A.~G. and {Jordan}, C.~A. and {Graham-Smith}, F. and {Espinoza}, C.~M. and {Stappers}, B.~W. and {Weltevrede}, P.},
        title = "{45 years of rotation of the Crab pulsar}",
      journal = {\mnras},
         year = 2015,
        month = jan,
       volume = {446},
       number = {1},
        pages = {857-864},
          doi = {10.1093/mnras/stu2118},
archivePrefix = {arXiv},
       eprint = {1410.0886},
 primaryClass = {astro-ph.HE},
       adsurl = {https://scixplorer.org/abs/2015MNRAS.446..857L}
}

@article{ocker_ne2025_2026,
  title = {{NE2025}: {An} {Updated} {Electron} {Density} {Model} for the {Galactic} {Interstellar} {Medium}},
  volume = {1002},
  issn = {0004-637X, 1538-4357},
  shorttitle = {{NE2025}},
  url = {http://arxiv.org/abs/2602.11838},
  doi = {10.3847/1538-4357/ae5825},
  number = {1},
  urldate = {2026-05-26},
  journal = {The Astrophysical Journal},
  author = {Ocker, S. K. and Cordes, J. M.},
  month = may,
  year = {2026},
  note = {arXiv:2602.11838 [astro-ph.GA]},
  pages = {3},
}

@software{newville_2025_16175987,
  author       = {Newville, Matthew and
                  Otten, Renee and
                  Nelson, Andrew and
                  Stensitzki, Till and
                  Ingargiola, Antonino and
                  Allan, Daniel and
                  Fox, Austin and
                  Carter, Faustin and
                  Rawlik, Michal},
  title        = {LMFIT: Non-Linear Least-Squares Minimization and
                   Curve-Fitting for Python
                  },
  month        = jul,
  year         = 2025,
  publisher    = {Zenodo},
  version      = {1.3.4},
  doi          = {10.5281/zenodo.16175987},
  url          = {https://doi.org/10.5281/zenodo.16175987},
  swhid        = {swh:1:dir:76742b0e41b1d2bff5a3716dd2376531f3a21db8
                   ;origin=https://doi.org/10.5281/zenodo.598352;visi
                   t=swh:1:snp:89f98ce93be53a573d85de0145f9174c827b7e
                   92;anchor=swh:1:rel:9528e5133d2d78c4036335f023b212
                   487cf1b632;path=lmfit-lmfit-py-0566445
                  },
}

@techreport{Lynch2026GBTMemo324,
  author      = {Lynch, Ryan and Marganian, Paul and Turner, Jacob and Hise, Tyler and
                 Baudler, Wolfgang and Sizemore, Dane and Jennings, Ross and Brandt, Joe and
                 Clark, Chris and Morin, Kirstin},
  title       = {Commissioning Report for the {GBT} Cyclic Spectroscopy Backend System},
  institution = {Green Bank Observatory},
  type        = {{GBT} Memo},
  number      = {324},
  year        = {2026},
  month       = jun,
  url         = {https://library.nrao.edu/public/memos/gbt/GBT_324.pdf}
}

@article{van_haarlem_lofar_2013,
	title = {{LOFAR}: {The} {LOw}-{Frequency} {ARray}},
	volume = {556},
	issn = {0004-6361},
	shorttitle = {{LOFAR}},
	url = {https://ui.adsabs.harvard.edu/abs/2013A%26A...556A...2V/abstract},
	doi = {10.1051/0004-6361/201220873},
	language = {en},
	urldate = {2022-10-13},
	journal = {Astronomy \& Astrophysics, Volume 556, id.A2, 53 pp.},
	author = {van Haarlem, M. P. and Wise, M. W. and Gunst, A. W. and Heald, G. and McKean, J. P. and Hessels, J. W. T. and de Bruyn, A. G. and Nijboer, R. and Swinbank, J. and Fallows, R. and Brentjens, M. and Nelles, A. and Beck, R. and Falcke, H. and Fender, R. and Hörandel, J. and Koopmans, L. V. E. and Mann, G. and Miley, G. and Röttgering, H. and Stappers, B. W. and Wijers, R. a. M. J. and Zaroubi, S. and van den Akker, M. and Alexov, A. and Anderson, J. and Anderson, K. and van Ardenne, A. and Arts, M. and Asgekar, A. and Avruch, I. M. and Batejat, F. and Bähren, L. and Bell, M. E. and Bell, M. R. and van Bemmel, I. and Bennema, P. and Bentum, M. J. and Bernardi, G. and Best, P. and Bîrzan, L. and Bonafede, A. and Boonstra, A.-J. and Braun, R. and Bregman, J. and Breitling, F. and van de Brink, R. H. and Broderick, J. and Broekema, P. C. and Brouw, W. N. and Brüggen, M. and Butcher, H. R. and van Cappellen, W. and Ciardi, B. and Coenen, T. and Conway, J. and Coolen, A. and Corstanje, A. and Damstra, S. and Davies, O. and Deller, A. T. and Dettmar, R.-J. and van Diepen, G. and Dijkstra, K. and Donker, P. and Doorduin, A. and Dromer, J. and Drost, M. and van Duin, A. and Eislöffel, J. and van Enst, J. and Ferrari, C. and Frieswijk, W. and Gankema, H. and Garrett, M. A. and de Gasperin, F. and Gerbers, M. and de Geus, E. and Grießmeier, J.-M. and Grit, T. and Gruppen, P. and Hamaker, J. P. and Hassall, T. and Hoeft, M. and Holties, H. A. and Horneffer, A. and van der Horst, A. and van Houwelingen, A. and Huijgen, A. and Iacobelli, M. and Intema, H. and Jackson, N. and Jelic, V. and de Jong, A. and Juette, E. and Kant, D. and Karastergiou, A. and Koers, A. and Kollen, H. and Kondratiev, V. I. and Kooistra, E. and Koopman, Y. and Koster, A. and Kuniyoshi, M. and Kramer, M. and Kuper, G. and Lambropoulos, P. and Law, C. and van Leeuwen, J. and Lemaitre, J. and Loose, M. and Maat, P. and Macario, G. and Markoff, S. and Masters, J. and McFadden, R. A. and McKay-Bukowski, D. and Meijering, H. and Meulman, H. and Mevius, M. and Middelberg, E. and Millenaar, R. and Miller-Jones, J. C. A. and Mohan, R. N. and Mol, J. D. and Morawietz, J. and Morganti, R. and Mulcahy, D. D. and Mulder, E. and Munk, H. and Nieuwenhuis, L. and van Nieuwpoort, R. and Noordam, J. E. and Norden, M. and Noutsos, A. and Offringa, A. R. and Olofsson, H. and Omar, A. and Orrú, E. and Overeem, R. and Paas, H. and Pandey-Pommier, M. and Pandey, V. N. and Pizzo, R. and Polatidis, A. and Rafferty, D. and Rawlings, S. and Reich, W. and de Reijer, J.-P. and Reitsma, J. and Renting, G. A. and Riemers, P. and Rol, E. and Romein, J. W. and Roosjen, J. and Ruiter, M. and Scaife, A. and van der Schaaf, K. and Scheers, B. and Schellart, P. and Schoenmakers, A. and Schoonderbeek, G. and Serylak, M. and Shulevski, A. and Sluman, J. and Smirnov, O. and Sobey, C. and Spreeuw, H. and Steinmetz, M. and Sterks, C. G. M. and Stiepel, H.-J. and Stuurwold, K. and Tagger, M. and Tang, Y. and Tasse, C. and Thomas, I. and Thoudam, S. and Toribio, M. C. and van der Tol, B. and Usov, O. and van Veelen, M. and van der Veen, A.-J. and ter Veen, S. and Verbiest, J. P. W. and Vermeulen, R. and Vermaas, N. and Vocks, C. and Vogt, C. and de Vos, M. and van der Wal, E. and van Weeren, R. and Weggemans, H. and Weltevrede, P. and White, S. and Wijnholds, S. J. and Wilhelmsson, T. and Wucknitz, O. and Yatawatta, S. and Zarka, P. and Zensus, A. and van Zwieten, J.},
	month = aug,
	year = {2013},
	pages = {A2},
}

@article{murphy_first_2021,
	title = {First results from the {REAL}-time {Transient} {Acquisition} backend ({REALTA}) at the {Irish} {LOFAR} station},
	volume = {655},
	issn = {0004-6361},
	url = {https://ui.adsabs.harvard.edu/abs/2021A&A...655A..16M},
	doi = {10.1051/0004-6361/202140415},
	urldate = {2026-01-30},
	journal = {Astronomy and Astrophysics},
	publisher = {EDP},
	author = {Murphy, P. C. and Callanan, P. and McCauley, J. and McKenna, D. J. and Fionnagáin, D. Ó. and Louis, C. K. and Redman, M. P. and Cañizares, L. A. and Carley, E. P. and Maloney, S. A. and Coghlan, B. and Daly, M. and Scully, J. and Dooley, J. and Gajjar, V. and Giese, C. and Brennan, A. and Keane, E. F. and Maguire, C. A. and Quinn, J. and Mooney, S. and Ryan, A. M. and Walsh, J. and Jackman, C. M. and Golden, A. and Ray, T. P. and Doyle, J. G. and Rigney, J. and Burton, M. and Gallagher, P. T.},
	month = nov,
	year = {2021},
	note = {ADS Bibcode: 2021A\&A...655A..16M},
	pages = {A16},
}

@ARTICLE{David_JOSS,
       author = {{McKenna}, D. J. and {Keane}, E. F. and {Gallagher}, P. T. and {McAuley}, J.},
        title ="{udpPacketManager: An International LOFAR Station Data (Pre-)Processor}",
      journal = {The Journal of Open Source Software},
         year = 2023
}

@article{men_transientx_2024,
	title = {{TransientX}: {A} high-performance single-pulse search package},
	volume = {683},
	issn = {0004-6361},
	shorttitle = {{TransientX}},
	url = {https://ui.adsabs.harvard.edu/abs/2024A&A...683A.183M},
	doi = {10.1051/0004-6361/202348247},
	urldate = {2025-05-22},
	journal = {Astronomy and Astrophysics},
	author = {Men, Yunpeng and Barr, Ewan},
	month = mar,
	year = {2024},
	note = {ADS Bibcode: 2024A\&A...683A.183M},
	pages = {A183},
}
\bibliographystyle{aasjournal}

\newpage
\appendix
\section{Burst Details per Epoch}

\begingroup
\tabletypesize{\scriptsize}
\LongTables 
\begin{deluxetable}{cccccccccccc}
\tablecaption{Burst and scattering statistics for all observations.\label{tab:bursts}}
\tablewidth{0pt}

\tablehead{
\colhead{Date} &
\colhead{$T_{\mathrm{obs}}$ (min)} &
\colhead{Bursts $(n)$} &
\colhead{GP Bursts} &
\colhead{$\mathrm{SNR}_{\mathrm{max}}$} &
\colhead{$\mathrm{DM}_{\mathrm{mean}}$} &
\colhead{$\mathrm{DM}_{\sigma}$} &
\colhead{$W_{\mathrm{mean}}$} &
\colhead{$W_{\sigma}$} &
\colhead{$\tau_{\mathrm{mean}}$ (ms)} &
\colhead{$\theta_{\mathrm{sun}}$ (deg)} &
\colhead{Burst Rate $(\mathrm{hr}^{-1})$}
}

\startdata
2020-03-17 & 476.0 & 21894 & 10841 & 2094.7 & 56.7691 & 0.0920 & 23.2 & 13.9 & 13.24 & 86.6 & 2760 \\
2020-06-26 & 946.2 & 13519 & 7156 & 2516.2 & 56.7649 & 0.1360 & 24.3 & 17.4 & 13.00 & 11.1 & 857 \\
2020-06-30 & 667.8 & 37176 & 19835 & 1720.1 & 56.7608 & 0.1494 & 24.8 & 17.5 & 12.94 & 14.9 & 3340 \\
2020-07-01 & 779.9 & 6988 & 3775 & 3601.0 & 56.7660 & 0.1405 & 24.2 & 17.3 & 12.31 & 15.9 & 538 \\
2020-07-07 & 80.3 & 3612 & 1530 & 2702.8 & 56.7339 & 0.1947 & 32.1 & 17.7 & 12.85 & 21.6 & 2700 \\
2020-07-14 & 35.5 & 1644 & 818 & 1116.1 & 56.7753 & 0.1065 & 23.3 & 15.6 & 11.23 & 28.2 & 2779 \\
2020-07-21 & 618.2 & 15171 & 8239 & 3062.5 & 56.7844 & 0.1113 & 24.5 & 15.3 & 13.47 & 34.9 & 1472 \\
2020-07-22 & 718.4 & 43300 & 22423 & 2559.6 & 56.7759 & 0.1361 & 26.5 & 16.1 & 13.33 & 35.9 & 3616 \\
2020-08-11 & 114.0 & 6489 & 3562 & 4107.6 & 56.7842 & 0.1055 & 25.1 & 14.6 & 13.80 & 54.9 & 3415 \\
2020-08-12 & 240.0 & 14836 & 7962 & 1545.2 & 56.7852 & 0.1055 & 25.1 & 14.7 & 13.66 & 55.7 & 3709 \\
2020-08-18 & 240.0 & 14298 & 7700 & 6597.7 & 56.7883 & 0.1130 & 27.3 & 14.7 & 13.85 & 61.5 & 3574 \\
2020-09-16 & 540.0 & 30503 & 16151 & 2077.6 & 56.7928 & 0.1086 & 25.7 & 14.6 & 14.01 & 89.6 & 3389 \\
2020-10-21 & 116.0 & 4913 & 2494 & 2363.9 & 56.7879 & 0.1232 & 33.7 & 12.6 & 22.16 & 123.8 & 2541 \\
2020-10-22 & 236.0 & 9854 & 5056 & 1860.5 & 56.7926 & 0.1142 & 33.9 & 12.6 & 22.78 & 124.9 & 2505 \\
2020-10-28 & 180.0 & 6512 & 3286 & 2165.5 & 56.7941 & 0.1254 & 36.4 & 12.9 & 24.18 & 130.8 & 2171 \\
2020-11-11 & 60.0 & 1951 & 867 & 206.6 & 56.7712 & 0.2102 & 44.8 & 8.8 & 26.09 & 144.8 & 1951 \\
2020-12-09 & 236.0 & 5711 & 2746 & 2253.9 & 56.8129 & 0.1385 & 43.5 & 8.8 & 29.02 & 173.0 & 1452 \\
2020-12-16 & 216.0 & 5759 & 2847 & 1970.6 & 56.8211 & 0.1358 & 43.5 & 8.4 & 28.31 & 178.7 & 1600 \\
2020-12-22 & 30.0 & 1697 & 803 & 557.4 & 56.8122 & 0.1576 & 43.2 & 8.7 & 35.58 & 172.6 & 3254 \\
2020-12-23 & 148.0 & 3232 & 1528 & 862.6 & 56.8149 & 0.1412 & 43.0 & 8.6 & 30.63 & 172.6 & 1310 \\
2020-12-29 & 60 & 3455 & 1581 & 433.8 & 56.7920 & 0.1648 & 42.4 & 9.4 & 36.19 & 165.6 & 3455 \\
2020-12-30 & 118.0 & 1770 & 789 & 827.5 & 56.7976 & 0.1527 & 41.0 & 10.1 & 30.81 & 165.6 & 900 \\
2021-01-05 & 358.0 & 10385 & 4859 & 1060.9 & 56.7982 & 0.1338 & 37.94 & 12.43 & 26.05 & 158.47 & 1741 \\
2021-01-12 & 129.3 & 3876 & 1838 & 758.3 & 56.7869 & 0.1342 & 34.8 & 14.3 & 22.86 & 151.4 & 1799 \\
2021-01-19 & 128.0 & 4291 & 2068 & 2351.1 & 56.7790 & 0.1451 & 35.8 & 14.1 & 20.99 & 144.3 & 2012 \\
2021-01-26 & 178.0 & 5983 & 2831 & 7214.0 & 56.7695 & 0.1459 & 36.5 & 13.4 & 21.48 & 137.2 & 2017 \\
2021-02-02 & 118.0 & 4287 & 2059 & 1804.4 & 56.7550 & 0.1700 & 38.4 & 13.2 & 22.71 & 130.1 & 2180 \\
2021-02-09 & 118.0 & 4359 & 2152 & 1321.8 & 56.7685 & 0.1260 & 34.4 & 14.2 & 22.25 & 123.0 & 2216 \\
2021-02-16 & 118.0 & 4045 & 1838 & 503.4 & 56.7454 & 0.1752 & 37.4 & 14.5 & 25.02 & 116.0 & 2057 \\
2021-02-23 & 118.0 & 4043 & 1941 & 1094.2 & 56.7661 & 0.1296 & 36.6 & 13.6 & 22.96 & 109.0 & 2056 \\
2021-03-03 & 118.0 & 3390 & 1682 & 752.8 & 56.7693 & 0.1210 & 37.9 & 12.8 & 25.62 & 100.9 & 1724 \\
2021-03-09 & 116.0 & 2833 & 1340 & 1321.2 & 56.7696 & 0.1189 & 36.4 & 13.0 & 24.68 & 94.9 & 1465 \\
2021-03-16 & 118.0 & 2594 & 1194 & 667.5 & 56.7712 & 0.1310 & 35.6 & 13.6 & 22.20 & 88.0 & 1319 \\
2021-03-17 & 117.2 & 2709 & 1236 & 403.8 & 56.7645 & 0.1508 & 36.3 & 13.5 & 23.68 & 87.0 & 1387 \\
2021-03-23 & 118.0 & 2660 & 1121 & 2813.0 & 56.7753 & 0.1531 & 38.7 & 13.0 & 25.80 & 81.0 & 1353 \\
2021-03-30 & 118.0 & 1962 & 863 & 352.9 & 56.7961 & 0.1337 & 41.8 & 10.8 & 24.33 & 74.2 & 998 \\
2021-04-06 & 118.0 & 1880 & 877 & 454.7 & 56.8077 & 0.1311 & 43.7 & 8.4 & 24.78 & 67.3 & 956 \\
2021-04-12 & 118.0 & 1962 & 913 & 965.7 & 56.8087 & 0.1412 & 44.4 & 7.4 & 33.11 & 61.4 & 998 \\
2021-04-20 & 118.0 & 2000 & 936 & 842.9 & 56.8007 & 0.1513 & 44.2 & 8.4 & 36.91 & 53.6 & 1017 \\
2021-04-27 & 118.0 & 2252 & 1012 & 1640.1 & 56.7842 & 0.1681 & 43.7 & 9.4 & 25.77 & 46.8 & 1145 \\
2021-05-05 & 162.0 & 2644 & 1185 & 563.7 & 56.7962 & 0.1454 & 42.4 & 9.8 & 34.47 & 39.1 & 979 \\
2021-05-12 & 118.0 & 2302 & 998 & 1147.0 & 56.7792 & 0.1681 & 42.7 & 9.4 & 29.75 & 32.3 & 1170 \\
2021-05-17 & 120.0 & 2232 & 1030 & 786.9 & 56.7859 & 0.1492 & 42.6 & 9.4 & 30.33 & 27.5 & 1116 \\
2021-05-25 & 118.0 & 3139 & 1238 & 1256.3 & 56.7208 & 0.2332 & 43.8 & 9.9 & 28.17 & 19.8 & 1596 \\
2021-06-02 & 118.0 & 2885 & 1274 & 221.9 & 56.7574 & 0.1774 & 38.8 & 12.8 & 14.74 & 12.2 & 1467 \\
2021-06-09 & 118.0 & 3586 & 1539 & 2648.9 & 56.7399 & 0.1972 & 37.6 & 13.7 & 18.42 & 5.6 & 1823 \\
2021-06-15 & 118.0 & 95 & 47 & 136.9 & 56.7711 & 0.1185 & 29.4 & 12.4 & 24.61 & 1.3 & 48 \\
2021-06-16 & 92.3 & 1577 & 465 & 277.2 & 56.6630 & 0.2646 & 37.3 & 15.1 & 20.50 & 1.7 & 1025 \\
2021-06-17 & 416.0 & 6791 & 3436 & 1908.8 & 56.5955 & 0.3566 & 47.4 & 7.3 & 15.43 & 2.6 & 980 \\
2021-06-23 & 118.0 & 4342 & 2092 & 878.9 & 56.7699 & 0.1219 & 27.4 & 15.2 & 15.41 & 8.0 & 2208 \\
2021-07-06 & 118.0 & 5704 & 2989 & 1259.1 & 56.7617 & 0.1337 & 26.5 & 16.0 & 12.21 & 20.2 & 2900 \\
2021-07-14 & 118.0 & 6834 & 3711 & 963.7 & 56.7634 & 0.1248 & 24.9 & 15.8 & 11.41 & 27.9 & 3475 \\
2021-07-22 & 118.0 & 6852 & 3745 & 1213.1 & 56.7649 & 0.1090 & 24.5 & 14.9 & 11.79 & 35.4 & 3484 \\
2022-03-03 & 58.0 & 841 & 343 & 792.4 & 56.7517 & 0.1620 & 43.5 & 8.4 & 25.22 & 101.2 & 870 \\
2022-03-09 & 58.0 & 475 & 195 & 225.4 & 56.7695 & 0.1321 & 43.6 & 8.1 & 26.71 & 95.2 & 491 \\
2022-03-16 & 29.0 & 134 & 51 & 666.1 & 56.7241 & 0.2095 & 42.7 & 10.1 & 23.43 & 88.4 & 277 \\
2022-03-22 & 58.0 & 366 & 169 & 88.5 & 56.7467 & 0.1571 & 37.0 & 12.2 & - & 82.3 & 379 \\
2022-03-29 & 48.0 & 258 & 119 & 611.5 & 56.7705 & 0.1133 & 34.7 & 11.3 & 22.69 & 75.3 & 322 \\
2022-04-12 & 58.0 & 479 & 225 & 306.9 & 56.7544 & 0.0809 & 24.3 & 16.6 & 9.14 & 61.6 & 496 \\
2022-04-26 & 58.0 & 1347 & 633 & 1394.8 & 56.7485 & 0.0675 & 13.1 & 10.9 & 6.67 & 48.0 & 1394 \\
2022-05-09 & 58.0 & 2430 & 1207 & 1203.2 & 56.7457 & 0.0756 & 16.0 & 12.2 & 9.22 & 35.4 & 2514 \\
2022-05-31 & 72.0 & 3887 & 1970 & 1954.6 & 56.7401 & 0.1060 & 20.7 & 15.2 & 8.91 & 14.4 & 3239 \\
2022-06-07 & 118.0 & 5893 & 3000 & 1303.7 & 56.7465 & 0.0888 & 21.3 & 14.9 & 9.69 & 7.7 & 2996 \\
2022-06-08 & 118.0 & 5517 & 2818 & 2667.6 & 56.7485 & 0.0822 & 21.2 & 14.6 & 9.52 & 6.8 & 2805 \\
2022-06-09 & 118.0 & 5659 & 2808 & 5408.1 & 56.7463 & 0.0841 & 21.1 & 14.8 & 9.20 & 5.8 & 2878 \\
2022-06-14 & 238.0 & 2337 & 455 & 146.4 & 56.5495 & 0.3290 & 45.4 & 10.1 & 12.92 & 1.6 & 589 \\
2022-06-15 & 478.0 & 3979 & 2079 & 83.8 & 56.4883 & 0.3687 & 49.1 & 0.7 & - & 1.3 & 500 \\
2022-06-17 & 223.5 & 7602 & 2873 & 1367.3 & 56.6700 & 0.2496 & 37.4 & 16.4 & 10.16 & 2.3 & 2040 \\
2022-06-21 & 59.0 & 1709 & 1083 & 735.0 & 56.5049 & 0.3969 & 48.7 & 3.6 & 10.30 & 5.9 & 1738 \\
2022-06-22 & 59.0 & 3340 & 1644 & 477.1 & 56.6514 & 0.2778 & 42.4 & 12.9 & 9.58 & 6.8 & 3397 \\
2022-06-23 & 59.0 & 4589 & 2206 & 981.9 & 56.7498 & 0.0948 & 23.7 & 14.6 & 9.57 & 7.7 & 4667 \\
2022-06-28 & 29.0 & 2560 & 1227 & 1676.2 & 56.7354 & 0.1328 & 24.4 & 15.8 & 9.24 & 12.4 & 5297 \\
2022-07-05 & 29.0 & 2189 & 1045 & 1306.2 & 56.7367 & 0.1221 & 23.1 & 15.3 & 8.84 & 19.1 & 4529 \\
2022-07-06 & 7.2 & 82 & 45 & 31.2 & 56.4201 & 0.3925 & 49.1 & 0.0 & - & 20.0 & 681 \\
2022-07-19 & 29.0 & 1883 & 863 & 580.2 & 56.7478 & 0.1065 & 25.2 & 14.0 & 10.74 & 32.4 & 3896 \\
2022-08-02 & 29.0 & 1775 & 875 & 1261.2 & 56.7505 & 0.1076 & 25.4 & 13.0 & 11.59 & 45.7 & 3672 \\
2022-08-09 & 29.0 & 1557 & 744 & 386.5 & 56.7538 & 0.0909 & 25.1 & 12.9 & 12.46 & 52.4 & 3222 \\
2022-08-16 & 58.0 & 1510 & 669 & 426.6 & 56.7474 & 0.1293 & 27.3 & 13.9 & 12.09 & 59.2 & 1562 \\
2022-08-26 & 29.0 & 1470 & 709 & 802.6 & 56.7559 & 0.0881 & 27.4 & 13.0 & 13.97 & 68.7 & 3042 \\
2022-09-23 & 29.0 & 1920 & 918 & 1143.7 & 56.7567 & 0.0937 & 26.2 & 13.1 & 14.97 & 95.8 & 3973 \\
2023-05-30 & 58.0 & 6762 & 3714 & 1188.1 & 56.7531 & 0.1086 & 21.2 & 14.1 & 8.28 & 15.6 & 6995 \\
2023-05-31 & 58.0 & 6396 & 3441 & 3306.0 & 56.7479 & 0.1213 & 23.6 & 15.0 & 8.25 & 14.6 & 6617 \\
2023-06-02 & 58.0 & 6204 & 3484 & 2751.2 & 56.7556 & 0.0986 & 22.1 & 13.6 & 8.40 & 12.7 & 6418 \\
2023-06-06 & 60.0 & 6224 & 3315 & 1278.2 & 56.7575 & 0.0990 & 22.5 & 13.4 & 8.77 & 8.9 & 6224 \\
2023-06-07 & 118.0 & 11435 & 5963 & 1013.0 & 56.7596 & 0.0893 & 22.7 & 13.5 & 8.90 & 7.9 & 5814 \\
2023-06-12 & 110.9 & 6740 & 2715 & 1155.4 & 56.6895 & 0.2217 & 34.1 & 16.6 & 9.54 & 3.3 & 3646 \\
2023-06-13 & 287.3 & 4749 & 2065 & 2061.1 & 56.7451 & 0.1336 & 24.4 & 15.8 & 11.54 & 2.5 & 992 \\
2023-06-27 & 58.0 & 4795 & 2530 & 685.7 & 56.7597 & 0.1228 & 23.4 & 15.0 & 9.76 & 11.3 & 4960 \\
2023-07-04 & 59.0 & 5797 & 2764 & 531.0 & 56.7439 & 0.1653 & 26.7 & 16.5 & 10.73 & 17.9 & 5895 \\
2023-07-11 & 118.0 & 9842 & 4384 & 1956.9 & 56.7397 & 0.1858 & 28.8 & 17.3 & 11.24 & 24.5 & 5004 \\
2023-07-18 & 58.0 & 4982 & 2614 & 1021.8 & 56.7700 & 0.1323 & 24.0 & 14.8 & 10.71 & 31.2 & 5154 \\
2023-07-25 & 58.0 & 2721 & 1480 & 1072.3 & 56.7739 & 0.1088 & 25.3 & 14.2 & 12.27 & 37.8 & 2815 \\
2023-08-01 & 58.0 & 3190 & 1652 & 1206.7 & 56.7797 & 0.1037 & 28.1 & 14.4 & 13.62 & 44.5 & 3300 \\
2023-08-08 & 58.0 & 2643 & 1349 & 2283.7 & 56.7779 & 0.1165 & 29.4 & 14.2 & 13.98 & 51.2 & 2734 \\
2023-08-15 & 58.0 & 2969 & 1551 & 1224.7 & 56.7802 & 0.1191 & 29.5 & 14.2 & 14.74 & 57.9 & 3072 \\
2023-08-29 & 60.0 & 2448 & 1219 & 1518.1 & 56.7887 & 0.1089 & 26.9 & 14.6 & 14.03 & 71.3 & 2448 \\
2023-09-05 & 58.0 & 2279 & 1117 & 405.0 & 56.7881 & 0.1038 & 24.4 & 13.7 & 12.94 & 78.1 & 2358 \\
2023-09-13 & 68.0 & 2654 & 1212 & 1936.1 & 56.7590 & 0.1667 & 30.1 & 15.6 & 12.44 & 85.8 & 2342 \\
2023-10-11 & 56.0 & 1745 & 789 & 802.2 & 56.7764 & 0.0969 & 20.7 & 12.8 & 10.54 & 113.3 & 1870 \\
2023-10-17 & 118.0 & 5751 & 2881 & 6153.5 & 56.7771 & 0.0856 & 19.7 & 12.7 & 10.32 & 119.1 & 2924 \\
2023-10-18 & 58.0 & 2766 & 1375 & 1612.4 & 56.7733 & 0.0981 & 19.8 & 12.8 & 10.25 & 120.1 & 2862 \\
2024-04-12 & 118.0 & 7935 & 4143 & 2274.2 & 56.7444 & 0.1098 & 19.6 & 13.9 & 8.10 & 61.3 & 4035 \\
2024-04-23 & 38.0 & 2979 & 1725 & 1255.7 & 56.7562 & 0.0914 & 18.4 & 13.3 & 8.23 & 50.5 & 4704 \\
2024-04-30 & 118.0 & 9627 & 5351 & 982.1 & 56.7595 & 0.0825 & 16.0 & 12.6 & 8.19 & 43.7 & 4895 \\
2024-05-08 & 108.0 & 4786 & 1894 & 1500.4 & 56.6975 & 0.2212 & 31.4 & 18.3 & 7.69 & 36.0 & 2659 \\
2024-05-21 & 118.0 & 10062 & 5348 & 2222.9 & 56.7391 & 0.1378 & 16.4 & 13.5 & 6.12 & 23.5 & 5116 \\
2024-05-28 & 120.0 & 11186 & 6240 & 1436.5 & 56.7472 & 0.1104 & 13.4 & 10.8 & 5.86 & 16.8 & 5593 \\
2024-06-04 & 118.0 & 10888 & 5778 & 1506.6 & 56.7341 & 0.1409 & 17.9 & 16.2 & 6.04 & 10.1 & 5536 \\
2024-06-11 & 118.0 & 10316 & 5686 & 1653.5 & 56.7523 & 0.0691 & 11.0 & 9.4 & 5.77 & 3.6 & 5246 \\
2024-06-12 & 128.0 & 10453 & 5712 & 1760.8 & 56.7572 & 0.0492 & 11.0 & 8.9 & 5.69 & 2.7 & 4900 \\
2024-06-18 & 118.0 & 10425 & 5556 & 1692.3 & 56.7407 & 0.1186 & 15.3 & 14.7 & 5.58 & 3.5 & 5301 \\
2024-06-19 & 118.0 & 13030 & 7249 & 1558.0 & 56.7502 & 0.0967 & 13.5 & 13.3 & 5.43 & 4.5 & 6626 \\
2024-06-25 & 58.0 & 6544 & 3829 & 1969.3 & 56.7496 & 0.0956 & 13.0 & 12.7 & 5.75 & 10.1 & 6770 \\
2024-06-26 & 58.0 & 6289 & 3562 & 2052.1 & 56.7483 & 0.1016 & 14.2 & 14.1 & 5.39 & 11.0 & 6506 \\
2024-07-02 & 58.0 & 6877 & 4030 & 3280.4 & 56.7452 & 0.1118 & 12.5 & 12.0 & 6.07 & 16.7 & 7114 \\
2024-07-03 & 58.0 & 6704 & 3930 & 1082.1 & 56.7466 & 0.1046 & 12.3 & 11.8 & 5.97 & 17.6 & 6935 \\
2024-07-09 & 38.0 & 4204 & 2466 & 1329.0 & 56.7472 & 0.1065 & 11.7 & 11.6 & 6.10 & 23.3 & 6638 \\
2024-07-10 & 38.0 & 4222 & 2487 & 1938.2 & 56.7473 & 0.1108 & 12.8 & 12.9 & 6.58 & 24.3 & 6667 \\
2024-07-16 & 38.0 & 3914 & 2173 & 1449.7 & 56.7499 & 0.0977 & 12.0 & 11.4 & 6.28 & 30.0 & 6180 \\
2024-07-17 & 38.0 & 3644 & 2083 & 1767.8 & 56.7457 & 0.1217 & 15.1 & 14.9 & 6.57 & 30.9 & 5754 \\
2024-07-23 & 238.0 & 28973 & 16628 & 2328.9 & 56.7483 & 0.1027 & 11.9 & 11.8 & 6.57 & 36.7 & 7304 \\
2024-07-24 & 318.0 & 38867 & 22347 & 2729.0 & 56.7488 & 0.0967 & 11.8 & 11.8 & 6.51 & 37.6 & 7334 \\
2024-08-06 & 58.0 & 5619 & 3275 & 1661.6 & 56.7601 & 0.0724 & 13.8 & 11.5 & 7.45 & 50.0 & 5813 \\
2024-08-07 & 58.0 & 5414 & 3105 & 1131.1 & 56.7600 & 0.0763 & 14.1 & 11.8 & 7.65 & 50.9 & 5601 \\
2024-08-12 & 88.0 & 7364 & 4047 & 1465.6 & 56.7543 & 0.1106 & 16.7 & 14.3 & 8.00 & 55.7 & 5021 \\
2024-08-13 & 58.0 & 4801 & 2737 & 1306.2 & 56.7627 & 0.0676 & 13.8 & 11.0 & 7.63 & 56.7 & 4967 \\
2024-08-14 & 58.0 & 4739 & 2660 & 1227.7 & 56.7625 & 0.0669 & 13.7 & 11.1 & 7.98 & 57.6 & 4903 \\
2024-08-20 & 58.0 & 3989 & 2119 & 1161.7 & 56.7671 & 0.0676 & 13.7 & 10.7 & 7.60 & 63.4 & 4127 \\
2024-08-27 & 58.0 & 3790 & 2004 & 1411.0 & 56.7647 & 0.0752 & 13.1 & 10.7 & 6.89 & 70.1 & 3921 \\
2024-09-03 & 58.0 & 4018 & 2119 & 1132.1 & 56.7717 & 0.0538 & 11.9 & 10.2 & 6.98 & 76.9 & 4157 \\
2024-09-10 & 16.6 & 1055 & 577 & 593.8 & 56.7708 & 0.0659 & 13.9 & 10.6 & 6.45 & 83.6 & 3802 \\
2024-09-14 & 58.0 & 3873 & 2134 & 1561.5 & 56.7689 & 0.0830 & 16.8 & 12.3 & 7.29 & 87.5 & 4007 \\
2024-09-17 & 58.0 & 4014 & 2164 & 1287.9 & 56.7695 & 0.0788 & 17.1 & 11.7 & 7.11 & 90.4 & 4153 \\
2024-10-08 & 58.0 & 4571 & 2366 & 1378.2 & 56.7474 & 0.1210 & 18.4 & 16.5 & 6.85 & 111.0 & 4729 \\
2024-10-10 & 58.0 & 4984 & 2857 & 2028.0 & 56.7612 & 0.0633 & 13.1 & 11.2 & 7.33 & 113.0 & 5156 \\
2024-10-21 & 58.0 & 3433 & 1554 & 1582.6 & 56.7291 & 0.1656 & 24.6 & 19.4 & 7.35 & 123.8 & 3552 \\
2024-10-22 & 58.0 & 4828 & 2598 & 948.5 & 56.7497 & 0.0827 & 13.4 & 12.8 & 7.17 & 124.8 & 4995 \\
2024-11-10 & 178.0 & 23098 & 13783 & 3224.9 & 56.7503 & 0.0656 & 12.2 & 10.8 & 5.63 & 143.8 & 7786 \\
2026-01-31 & 216.3 & 17384 & 9458 & 1699.8 & 56.7310 & 0.0655 & 12.3 & 10.1 & 6.40 & 132.3 & 4822 \\
2026-02-01 & 240.0 & 20817 & 11338 & 1448.2 & 56.7312 & 0.0659 & 12.4 & 10.4 & 6.47 & 131.4 & 5204 \\
2026-02-02 & 120.0 & 20652 & 11334 & 1839.1 & 56.7317 & 0.0645 & 12.4 & 10.3 & 6.55 & 130.4 & 10326 \\
2026-02-03 & 240.0 & 20976 & 11634 & 1453.8 & 56.7328 & 0.0602 & 12.2 & 9.7 & 6.46 & 129.4 & 5244 \\
2026-02-04 & 240.0 & 20727 & 11177 & 1406.2 & 56.7310 & 0.0735 & 13.1 & 11.0 & 6.44 & 128.4 & 5182 \\
2026-02-23 & 177.0 & 9074 & 4904 & 1046.7 & 56.7313 & 0.0699 & 13.2 & 10.9 & 7.12 & 109.2 & 3077 \\
2026-03-02 & 120.0 & 8792 & 4721 & 1174.9 & 56.7318 & 0.0710 & 13.2 & 10.9 & 6.94 & 102.3 & 4396 \\
2026-03-06 & 120.0 & 8694 & 4372 & 1071.1 & 56.7227 & 0.1191 & 18.1 & 15.4 & 6.84 & 98.3 & 4347 \\
2026-03-09 & 240.0 & 18501 & 9977 & 1166.0 & 56.7356 & 0.0580 & 12.8 & 9.7 & 6.54 & 95.2 & 4625 \\
2026-03-10 & 118.0 & 8875 & 4631 & 1792.1 & 56.7277 & 0.1011 & 16.1 & 13.6 & 6.63 & 94.3 & 4513 \\
2026-03-11 & 236.0 & 18312 & 9892 & 100.0 & 56.7347 & 0.0640 & 12.7 & 9.9 & - & 93.2 & 4656 \\
2026-03-12 & 118.0 & 9016 & 4908 & 1355.7 & 56.7356 & 0.0544 & 12.5 & 9.5 & 6.40 & 92.2 & 4584 \\
2026-03-13 & 177.0 & 9081 & 4880 & 1114.3 & 56.7339 & 0.0620 & 12.7 & 10.2 & 6.65 & 91.2 & 3078 \\
2026-03-17 & 236.0 & 18064 & 9818 & 1151.1 & 56.7339 & 0.0582 & 11.9 & 9.5 & 6.41 & 87.2 & 4593 \\
\enddata

\end{deluxetable}
\endgroup

\vspace{10pt} 

\end{document}